\documentclass[11pt]{article}
\usepackage[utf8]{inputenc}
\usepackage{fullpage}
\usepackage[square,sort,comma,numbers]{natbib}
\usepackage{amsmath}
\usepackage{hyperref}
\usepackage{graphicx}
\usepackage{setspace}
\usepackage{bm} 
\usepackage{amssymb}
\usepackage{subcaption}
\usepackage{todonotes}
\usepackage{xcolor}

\DeclareMathOperator{\var}{var}

\title{A fully Bayesian sparse polynomial chaos expansion approach with joint priors on the coefficients and global selection of terms}
\author{Paul-Christian Bürkner$^{1*}$, Ilja Kröker$^{1,2}$, Sergey Oladyshkin$^{1,2}$, Wolfgang Nowak$^{1,2}$}
\date{
  $^1$ Stuttgart Center for Simulation Science, University of Stuttgart, Germany\break 
  $^2$ Department of Stochastic Simulation and Safety Research for Hydrosystems, Institute for Modelling Hydraulic and Environmental Systems, University of Stuttgart, Germany \break
  $^*$ Corresponding author; Email: paul.buerkner@gmail.com; Address: Universitätsstr. 32, 70569 Stuttgart, Germany
}

\begin{document}

\maketitle


\newpage
\begin{abstract}
Polynomial chaos expansion (PCE) is a versatile tool widely used in uncertainty quantification and machine learning, but its successful application depends strongly on the accuracy and reliability of the resulting PCE-based response surface. High accuracy typically requires high polynomial degrees, demanding many training points especially in high-dimensional problems through the curse of dimensionality. So-called sparse PCE concepts work with a much smaller selection of basis polynomials compared to conventional PCE approaches and can overcome the curse of dimensionality very efficiently, but have to pay specific attention to their strategies of choosing training points. Furthermore, the approximation error resembles an uncertainty that most existing PCE-based methods do not estimate. In this study, we develop and evaluate a fully Bayesian approach to establish the PCE representation via joint shrinkage priors and Markov chain Monte Carlo. The suggested Bayesian PCE model directly aims to solve the two challenges named above: achieving a sparse PCE representation and estimating uncertainty of the PCE itself. The embedded Bayesian regularizing via the joint shrinkage prior allows using higher polynomial degrees for given training points due to its ability to handle underdetermined situations, where the number of considered PCE coefficients could be much larger than the number of available training points. We also explore multiple variable selection methods to construct sparse PCE expansions based on the established Bayesian representations, while globally selecting the most meaningful orthonormal polynomials given the available training data. We demonstrate the advantages of our Bayesian PCE and the corresponding sparsity-inducing methods on several benchmarks.


\end{abstract}


\newpage
\section{Introduction}
The polynomial chaos expansion (PCE) was originally introduced through the theory of homogeneous chaos by Norbert Wiener \cite{Wiener1938} in 1938.  
Convergence in the $L^2$-sense was shown by Cameron and Martin \cite{MR3364576}. Since Ghanem and Spanos \cite{Ghanem_Spanos_1991_SFEM_book} popularized PCE-based stochastic Galerkin techniques in applications, many authors have investigated the construction \cite{OladNowak_RESS2012,xiu2002wiener} 
and limitations of PCE \cite{MR2855645,oladyshkin2018incomplete}
and applied PCE techniques to many application problems \cite{MR2501693,MR2121216,koppel2019comparison,sudret2008,witteveen2006modeling,Xiu2003}. Most applications use the PCE as a response surface technique to provide surrogates for computationally expensive models, which can then be used for uncertainty quantification and related tasks.
Beyond fundamental properties like fast evaluation and regularity
of PCE surrogates, the PCE can provide stochastic moments of the approximated model response (like mean and variance) and global sensitivity indices \cite{sudret2008} in analytical form, i.e. without further evaluations of the original model.

Despite these advantages, the feasibility and efficiency of PCE approximations tend to suffer because the number of expansion terms and the corresponding required training runs of the original model grow fast with increasing input dimension (curse of dimensionality). To extend the practical applicability of PCE to higher-dimensional problems, several later techniques seek sparse PCE approximations, where only selected expansion terms have non-zero coefficients.
Available ideas include sparse quadrature and stochastic collocation \cite{doi:10.1137/120890715,Olad_al_CG2009,ZENG20121} or stratified sampling methods \cite{hadigol2018least,Keese2003}. In particular, there are sparse approaches based on classical regression \cite{ahlfeld2016samba,blatman2008sparse} and Bayesian sparse PCE approaches \cite{lusparse,mohammadi2022uncertaintyaware,pan2020,shao2017}. To the best of our knowledge, the existing Bayesian sparse PCE approaches see the PCE's expansion coefficients as Gaussian random variables that offers the pathway for Bayesian sparse learning as for relevance vector machine \cite{tipping2001sparse} to obtain the posterior distribution of the PCE coefficients.
Some real-world applications require regularization strategies on the PCE to obtain reliable surrogate models. Therefore, the PCE Ansatz was combined with several regularization strategies, beginning regularization via Moore–Penrose inverse \cite{moore1920reciprocal,penrose1956best} or $\ell_2$-norm regularized least-squares classical regularization in \cite{6425919}, up to Elastic Net \cite{WOS:000490766100036} and Entropy-based \cite{WOS:000742966300001} regularization concepts. Among these, the Elastic net idea demonstrates that regularization and sparsity can be addressed together, as Elastic net includes a sparsity-inducing $\ell^1$-term in its regularization.

In this paper, we focus on Bayesian PCE approaches as they can quantify the epistemic uncertainty of the PCE (remaining after training, not the uncertainty of model responses due to potentially uncertain parameters of the true model) through the implied posterior distribution of the PCE coefficients. However, to compute the posterior analytically, the corresponding prior distributions of the PCE coefficients have to be Gaussian, in which case the posterior is Gaussian as well \cite{BDA3} similarly to the existing Bayesian sparse PCEs \cite{lusparse,mohammadi2022uncertaintyaware,pan2020,shao2017}. This severely limits the flexibility of Bayesian PCEs. In particular, it is prone to overfitting (i.e., fitting too closely on the training data and thus generalizing poorly) in cases with few training points and correspondingly high number of considered PCE terms, as the implied ridge-regularization (purely $\ell^2$-based) is insufficient in such cases \cite{bernardo_shrink_2011}.

In the wider field of statistics, much more strongly regularizing (shrinkage) priors have been demonstrated to provide good approximation performance even in highly under-determined problems \cite{bhadra_default_2016, piironen_sparsity_2017, van_erp_shrinkage_2019,  zhang_bayesian_2020}, but so far have not found their way into the PCE literature. Due to complex geometry of the implied posteriors, sampling techniques (in particular Markov chain Monte-Carlo (MCMC)) are required to achieve trustworthy approximations of the posterior \cite{BDA3, hoffman2014}. Gaussian approximations are insufficient in these cases, as neither the marginal posterior distributions would be sufficiently Gaussian, nor would the joint distribution necessarily follow a Gaussian dependence structure \cite{piironen_sparsity_2017, yao_diditwork_2018, zhang_bayesian_2020}.

In this study, we develop and evaluate a new Bayesian sparse PCE approach. Our key idea is to use a joint shrinkage prior on the PCE coefficients, specifically the so-called R2D2 prior \cite{zhang_bayesian_2020} as summarized in Section~\ref{R2D2-prior}. The proposed R2D2-based PCE model aims to achieve both a sparse PCE representation and accurate uncertainty estimation of the PCE itself. The embedded Bayesian regularization via the R2D2 prior allows using higher polynomial degrees for given training points. It can even handle under-determined situations, where the number of considered PCE coefficients is much larger than the number of available training points. Moreover, our fully Bayesian approach estimates uncertainty of the PCE itself via MCMC estimation of the posterior. And, by doing so, it avoids Gaussian assumptions on both the posterior of the coefficients and the implied response surface. On top of that, we also explore multiple variable selection methods to construct sparse PCE expansions with our Bayesian approach, globally selecting the most meaningful polynomials given the available training data (see Section~\ref{variable-selection}).

We demonstrate the advantages of the R2D2-based Bayesian PCE and of the corresponding sparsity-inducing methods on several benchmarks in Section~\ref{case-studies}. In these benchmarks, we employ the arbitrary polynomial chaos (aPC) generalization of the conventional PCE. The aPC satisfies orthonormality in data-driven situations and so assures flexibility for various applications (see Section~\ref{general-PCE}). The test cases range from widespread low- to high-dimensional benchmark functions up to a simulation-based benchmark in the area of carbon dioxide storage in the subsurface.  

\section{Methods}

\subsection{Polynomial chaos expansion}
\label{general-PCE}

The Polynomial Chaos Expansion (PCE) \cite{Wiener1938} expresses the dependence of a model response (output) on a set of input parameters using a projection onto an orthonormal polynomial basis \cite{Ghanem_Spanos_1991_SFEM_book,MR3364576}. Formally, we will consider a univariate response $y = y(x,t)$ generated from some true -- but potentially unknown -- model $\mathcal{M}(x,t,\bm{\omega})$ that depends on space $x$, time $t$, and on a vector of uncertain input parameters $\bm{\omega}=\left\{\omega_1,...,\omega_N \right\}$ from the input space $\Omega$ with a given probability measure $\Gamma_{\omega}$.
In practice, we often do not know the exact analytical form of $\mathcal{M}(x,t,\bm{\omega})$ or cannot evaluate it due to computational challenges. Hence, we seek to find a surrogate model that approximates the true model as closely as possible while having some favourable practical or computational properties.  When using PCE as a surrogate \cite{MR0020230,Wiener1938}, we approximate the true model as
\begin{eqnarray}
\label{PCE}
  \mathcal{M}(x,t,\bm{\omega}) &\approx& \sum_{i=0}^{M} c_i(x,t) \Psi_i(\bm{\omega}),
\end{eqnarray}
where $\Psi_i(\bm{\omega})$ are basis functions of a multivariate polynomial basis $\{ \Psi_0(\bm{\omega}),..., \Psi_M(\bm{\omega})\}$ that is orthonormal in ($\Omega, \Gamma_{\omega})$, and $c_i = c_i(x, t)$ are corresponding coefficients that determine the form of the expansion in Equation (\ref{PCE}).
If the model $\mathcal{M}$ contains space or time coordinates $(x,t)$, then we consider the coefficients $c_i(x,t)$ as defined independently for each point in space and time. For easier notation, we will omit these coordinates from now on unless explicitly required. 

In the widespread total-degree truncation, the total number $M$ of expansion terms in Equation (\ref{PCE}) depends combinatorically on the number of input parameters $N$ and on the degree of expansion $d$ (i.e., maximal degree of considered polynomials) as 
\begin{eqnarray}
\label{PCE_size}
  M = \frac{(N+d)!}{N! \, d!}.
\end{eqnarray}

Assuming statistical independence of the input parameters, the corresponding multivariate polynomial basis $\{ \Psi_0(\bm{\omega}),..., \Psi_M(\bm{\omega})\}$ is the tensor product of univariate orthonormal polynomials $\{\phi^{(0)}_j$, $\ldots$, $\phi^{(d)}_j \}$ up to degree $d$ for each input parameter $\omega_j$:
\begin{eqnarray}
\label{basis_multi}
\Psi_{\alpha}(\bm{\omega})=\prod^N_{j=1} \phi^{(\alpha_j)}_j(\omega_j), \quad \sum^N_{j=1} \alpha_j \leq d,
\end{eqnarray}
where $\alpha$ is a multivariate index that contains the combinatoric information how to enumerate all possible products of individual univariate orthonormal polynomials $\{\phi^{(0)}_j$, $\ldots$, $\phi^{(d)}_j \}$ for each input parameter $\omega_j$. 

This set of univariate polynomials forms an orthonormal basis of polynomial degree $d$ for each input parameter $\omega_j$ ($j=1,.., N$):
\begin{eqnarray}
\label{multi_inde}
\int_{\Omega} \phi^{(p)}_j(\omega_j) \, \phi^{(q)}_j(\omega_j) \, d \, \Gamma(\omega_j) = \delta_{p,q}, \quad \forall p,q =0,...,d \
\end{eqnarray}
where $\delta_{p,q}$ represents the Kronecker delta.

There are several ways to construct the required univariate  polynomials. The original PCE theory \cite{Wiener1938} was developed for normally distributed input parameters $\omega_j$, leading to Hermite polynomials as basis. Later works (\cite{xiu2002wiener} and \cite{Xiu2003}) suggested the Askey scheme \cite{Askey1985} of polynomials that extend the original PCE theory to several other common parametric distributions (Gamma, Beta, Uniform, etc.). Like Hermite polynomials are optimal for the normal distribution, Legendre polynomials are optimal for the uniform distribution of inputs parameter $\omega_j$. Going beyond these distributions, univariate orthogonal polynomials can be also obtained using three-term recurrences \cite{favard1935polynomes}, Gram–Schmidt orthogonalization \cite{witteveen2006modeling}, the Stieltjes procedure \cite{stieltjes1884quelques} or by a data-driven approach \cite{OladNowak_RESS2012} know as arbitrary polynomial chaos (aPC). For a comprehensive discussion of the requirements for existence and completeness of orthonormal polynomial bases, we refer to \cite{MR2855645,MR2061539,MR3364576}.

Any of the above procedures can be used to construct the required univariate orthonormal polynomials. However, in many applications, exact knowledge of the involved probability distribution \cite{MR3364576} is not available, or would require additional assumptions \cite{RedHorse2004}. To assure highly parsimonious and, if necessary, purely data-driven representation, we will pay especial attention to the aPC generalization.  The aPC technique adapts to arbitrary probability distribution shapes of input parameters. Its univariate orthonormal basis can be inferred from limited input data through the $2d$ raw stochastic moments \cite{OladNowak_RESS2012} as follows: 
\begin{eqnarray}
\label{poly_monomials}
  \phi^{(\alpha_j)}_j(\omega_j) = \frac{1}{\sqrt{\kappa_{\alpha_j}}}\sum_{i=0}^{\alpha_j} m^{(\alpha_j)}_i \omega_j^i, \quad \alpha_j=0, \ldots, d,
\end{eqnarray}
where $\kappa_{\alpha_j}=m^{(\alpha_j)}_{\alpha_j}$ is a constant representing the norm of the univariate polynomial. The corresponding monomial coefficients $m^{(\alpha_j)}_i$ could be defined according to the raw moments of the inputs $\omega_j$ as follows \cite{OladNowak_RESS2012,oladyshkin2018incomplete}:
\begin{eqnarray}
\label{orth_matrix_dd}     
\begin{bmatrix}      
\mu_0(\omega_j) & \mu_1(\omega_j) & \ldots & \mu_{\alpha_j}(\omega_j) \cr           
\vdots & \vdots & \vdots & \vdots \cr            
\mu_{\alpha_j-1}(\omega_j) & \mu_{\alpha_j}(\omega_j) & \ldots & \mu_{2\alpha_j-1}(\omega_j) \cr
0 & 0 & \ldots & 1
\end{bmatrix} 
\begin{bmatrix}      
m^{(\alpha_j)}_0 \cr                
\vdots \cr            
m^{(\alpha_j)}_{\alpha_j-1} \cr                
m^{(\alpha_j)}_{\alpha_j} \cr    
\end{bmatrix} 
=
\begin{bmatrix}
0 \cr                
\vdots \cr            
0 \cr                
1 \cr    
\end{bmatrix},
\end{eqnarray}
where $\mu_{k}(\omega_j)$ denotes the $k$-th raw stochastic moment of the input $\omega_j$. The matrix on the left-hand side of Equation (\ref{orth_matrix_dd}) is known as the Hankel matrix of moments \cite{Karlin1968}, and its properties have been analysed in \cite{Lindsay1989}. The polynomials are real if, and only if, the Hankel matrix of moments is positive definite, see also the related Hamburger moment problem, e.g. \cite{akhiezer1965classical,MR2855645,shohat1943problem,MR3364576}. The aPC formulation offers a certain flexibility in  modelling, as the required statistical moments in the Hankel matrix can be evaluated directly from an available data set of observed input parameter values, but they could also be determined analytically from suitable, parametric distributions. For example, Equation (\ref{poly_monomials}) reconstructs the Legendre polynomials when inserting the raw moments $\mu_{k}(\omega_j)$ of the uniform distribution.

Regardless of the specific PCE technique, each expansion coefficient $c_{i}$ in Equation (\ref{PCE}) indicates how much variance one or another expansion term brings into the overall composition. Therefore, the PCE is closely related to global sensitivity analysis \cite{sudret2008,zhang2016evaluation}. Namely, due to the orthonormal representation, Sobol \cite{sobol1990sensitivity} sensitivity indices $S_{i}$ for each individual expansion term is directly given by \cite{oladyshkin2012global}: 
\begin{equation}
\label{sobol-index}
 S_{i}= \frac{c^2_{i}}{\sum_{i'=1}^{M}c_{i'}^2},
\end{equation}
where the Sobol index $S_{i}$ reflects the relative partial contribution of each single PCE term to the total variance of the response $y$.

\subsection{Conventional deterministic computation of expansion coefficients}
\label{deterministic-PCE}

Methods to compute the expansion coefficients $c_i$ in Equation (\ref{PCE}) can be sub-divided into intrusive and non-intrusive approaches. The intrusive approach (e.g. Galerkin projection) requires manipulation of the governing equations for stochastic analysis in application cases where (partial) differential equations are analyzed \cite{Ghanem_Spanos_1991_SFEM_book, koppel2019comparison}. Due to the necessity of symbolic manipulations, this procedure may become analytically cumbersome and transfer from one application to another requires addition efforts.  For that reason, non-intrusive approaches like sparse quadrature \cite{Keese2003}, the probabilistic collocation method \cite{Li2007,Olad_al_CG2009,Villadsen1978} or numerical regression \cite{beckers2020bayesian,scheurer2021surrogate} have received increasing attention for applied tasks, as they do not demand any modification of original model/simulation codes. For the sake of flexibility, we will choose non-intrusive approaches in our current work. 

The non-intrusive approaches employ so-called collocation points that could be seen as training points corresponding to sets of parameter values $\bm{\omega}$ for which the original model is run. Let us denote the training input parameters by $\bm{\Omega}_T=\{\bm{\omega}_{1},\ldots,\bm{\omega}_{T} \}$ and the corresponding training responses by $\mathbf{y}_{T} = \{y_{1}, \ldots, y_{T}\}$, where $T$ is the number of the training points. It is well known \cite{OladNowak_RESS2012} that the selection of training points strongly influences the performance of the overall PCE representation in Equation (\ref{PCE}). According to Gaussian integration theory \cite{Villadsen1978}, the optimal training points are the roots of the orthonormal polynomial one degree higher than the overall expansion degree $d$. The Gaussian integration strategy avoids additional oscillations known as the Runge phenomenon \cite{runge1901empirische}. However, our current work is not limited to any particular selection strategy of the training points and, hence, we will keep the formulation as general as possible.

Regardless of their design, the chosen training inputs $\bm{\Omega}_T$ and the corresponding training responses $\mathbf{y}_{T}$ can be inserted into Equation (\ref{PCE}) to obtain the following linear system of equations that determines the expansion coefficients $c_i$:
\begin{eqnarray}
\label{PCE_linsys}
\begin{bmatrix}      
\Psi_{0}(\bm{\omega}_1) & \ldots& \Psi_{M}(\bm{\omega}_1) \cr          
\vdots & \vdots & \vdots \cr            
\Psi_{0}(\bm{\omega}_T) & \ldots& \Psi_{M}(\bm{\omega}_T) \cr    
\end{bmatrix} 
\begin{bmatrix}      
c_1 \cr                
\vdots \cr            
c_M \cr    
\end{bmatrix} 
=
\begin{bmatrix}
y_{1} \cr                
\vdots \cr            
y_{T} \cr    
\end{bmatrix},
\end{eqnarray}
or in the following matrix form:
\begin{eqnarray}
\label{PCE_linsys_matrix}
  \bm{\Psi} \cdot \bm{c} = \mathbf{y}_{T},
\end{eqnarray}
where $\bm{\Psi}$ is the co-called design matrix of size $T \times M$ and $\bm{c}$ is the vector of size $M$ containing all unknown expansion coefficients. 

There are several deterministic ways to obtain the expansion coefficients $c_i$ from the linear system (\ref{PCE_linsys_matrix}). If the number $M$ of expansion terms is exactly equal to the number $T$ of training points, the system (\ref{PCE_linsys_matrix}) is fully determined, e.g., as in the probabilistic collocation method \cite{Li2007,Olad_al_CG2009,Villadsen1978}. Therefore, the expansion coefficients could be found by: 
\begin{eqnarray}
\label{PCE_solution_classical}
  \bm{c} = \bm{\Psi}^{-1}\cdot \mathbf{y}_{T},
\end{eqnarray}
When the number $T$ of training points is higher than the number $M$ of expansion terms, the linear system  (\ref{PCE_linsys_matrix}) is over-determined. Then, the expansion coefficients can be obtained by least-squares minimization \cite{hadigol2018least,mura2020least} via Moore-Penrose inversion \cite{moore1920reciprocal,penrose1956best} (also known as pseudoinverse) on the $T$ regression points: 
\begin{eqnarray}
\label{PCE_solution_ls}
  \bm{c} = (\bm{\Psi}^T\cdot \bm{\Psi})^{-1} \cdot \bm{\Psi}^T \cdot \mathbf{y}_{T}.
\end{eqnarray}
However, since the training points do not necessarily correspond to optimal Gaussian roots \cite{Villadsen1978} \cite{OladClasNow_CG2013,beckers2020bayesian}, additional  regularization \cite{tikhonov1977solutions} helps to mitigate the Runge phenomenon \cite{runge1901empirische}. With Gaussian regularization, Equation (\ref{PCE_linsys_matrix}) becomes
\begin{eqnarray}
\label{PCE_solution_pinv}
  \bm{c} = (\bm{\Psi}^T\cdot \bm{\Psi}+\lambda \bm{I})^{-1} \cdot \bm{\Psi}^T \cdot \mathbf{y}_{T},
\end{eqnarray}
where $\lambda$ is the regularization parameter and $\bm{I}$ is the identity matrix of appropriate dimension. Equation (\ref{PCE_solution_pinv}) is a generalized formulation of Equation (\ref{PCE_solution_ls}), and returns to the original least-squares formulation at $\lambda=0$. It also contains the fully-determined problem in Equation (\ref{PCE_solution_classical}) for $\lambda=0$ and $M=T$. Moreover, Equation (\ref{PCE_solution_pinv}) also works when the original linear system (\ref{PCE_linsys_matrix}) is under-determined, i.e., when $T<M$. For these reasons, Equation (\ref{PCE_solution_pinv}) has been used in practice \cite{Olad_EP2013,scheurer2021surrogate,papi2021uncertainty}, relying on a publicly available aPC code implementation \cite{aPC_Toolbox}. Alternative ideas to regularize the deterministic PCE coefficients could be used following the ideas in \cite{6425919,WOS:000742966300001}.

\subsection{Bayesian PCE}
\label{bayesian-PCE}

PCE can also be performed in a fully probabilistic (Bayesian) framework, by treating it as a Bayesian linear regression model with the evaluated polynomials as predictors \citep{shao2017}. For this purpose, we assume the training responses $\mathbf{y}_{T}$ to be normally distributed with mean vector $\mathbf{\mu}$ equal to the PCE approximation and residual variance $\sigma^2$. Together, this forms the likelihood $p(\mathbf{y}_{T} \,|\, \mathbf{c}, \sigma^2)$ of the data $\mathbf{y}_{T}$, given the PCE coefficients $\mathbf{c}$ and residual variance $\sigma^2$ as model parameters. The latter parameter captures both potential noise in the observed responses and approximation errors caused by the truncation of the PCE to a finite number of $M$ polynomials. When additionally considering prior distributions for all the parameters, the following Bayesian PCE model arises:
\begin{align}
\label{bayesianPCE}
  \mathbf{y}_{T} &\sim \text{normal}(\mathbf{\mu}, \sigma^2) \\
  \mathbf{\mu} &= \sum_{i=0}^{M} c_i \Psi_i(\bm{\omega}) \label{linpred} \\
  \mathbf{c} &\sim p(\mathbf{c}) \\
  \sigma^2 &\sim p(\sigma^2),
\end{align}
which implies the posterior distribution of the PCE parameters given the data:
\begin{equation}
\label{posterior}
    p(\mathbf{c}, \sigma^2  \,|\, \mathbf{y}_{T}) = \frac{p(\mathbf{y}_{T} \,|\, \mathbf{c}, \sigma^2) \, p(\mathbf{c}) \, p(\sigma^2)}{\int p(\mathbf{y}_{T} \,|\, \mathbf{c}, \sigma^2) \, p(\mathbf{c}) \, p(\sigma^2) \, d \mathbf{c} \, d \sigma^2}
\end{equation}

The specific properties of the posterior and hence of the Bayesian PCE solution will depend on the choice of priors $p(\mathbf{c})$ and $p(\sigma^2)$. For example, if all the priors were (improper) flat, $p(\mathbf{c}) \propto 1$ and $p(\sigma^2) \propto 1$, the posterior mode would be equal to the maximum likelihood estimate, which for the here considered PCE, is also equal to the least-squares solution \eqref{PCE_solution_classical}. If multivariate normal priors are used as $p(\mathbf{c})$ and inverse-gamma priors are used as $p(\sigma^2)$, the priors are conjugate to the likelihood \citep{BDA3}. This means that the posterior will be of the same functional form as the prior, with multivariate normal posterior for $\mathbf{c}$ (conditionally on $\sigma^2$) and inverse-gamma posterior for $\sigma^2$. If, for example, the prior mean and covariance matrix for $\mathbf{c}$ were $\mu_{\rm prior} = 0$ and $\Sigma_{\rm prior} = \frac{1}{\lambda} I$, respectively, with regularization parameter $\lambda$ and identity matrix $\bm{I}$, the posterior mean would be equal to the regularized PCE estimate \eqref{PCE_solution_pinv}. As such, there exist natural connections between classical and Bayesian PCE despite conceptually different approaches by which they are derived.

The posterior forms the main target of Bayesian inference but is, with the exception of conjugate cases, usually analytically intractable due to the intractability of the integral that occurs as denominator in \eqref{posterior}. In the current work we will go beyond the Gaussian assumption behind the prior and posterior of PCE coefficients that have been used in previous versions of Bayesian PCE approaches \cite{lusparse,mohammadi2022uncertaintyaware,pan2020,shao2017}. For general cases, we will have to resort to posterior approximation methods that only require evaluating the likelihood and prior. Out of all such approximation methods, MCMC is perhaps the most important class \citep{BDA3}. MCMC enables us to draw dependent random samples from the posterior, which can subsequently be used to compute relevant posterior quantities such as posterior means or variances. Their approximation errors approach zero for increasing number of samples under very general conditions, but for small number of samples or poor convergence, the approximation errors may be substantial \citep{BDA3, vehtari2021}. As such, in practice, convergence of the MCMC procedure and sufficient sampling efficiency is verified by both graphical and numerical diagnostics before estimates are computed \citep{gabry2019,vehtari2021}.

\subsection{The R2D2 prior}
\label{R2D2-prior}

We are specifically interested in scenarios with sparse training data (i.e., small number $T$ of training points), where we need to find sparse PCE solutions that include as few polynomials as possible without sacrificing too much approximation performance. Sparsity does not always imply that PCE coefficients of irrelevant polynomials are found to be exactly zero. Instead, approximately zero (\emph{weakly sparse}) will usually be good enough and is easier to achieve, since the underlying parameter space remains continuous rather than reducing to a point mass \citep{piironen_sparsity_2017}. Priors that induce weak sparsity by strong regularization of irrelevant (or little relevant) terms are known as \emph{shrinkage priors} \citep{piironen_sparsity_2017, van_erp_shrinkage_2019}. We will focus specifically on one of these priors, the $R^2$-\emph{D}irichlet-\emph{D}ecomposition (R2D2) prior \citep{zhang_bayesian_2020}, whose assumptions match very well with the needs of sparse PCE. The R2D2 prior belongs to the class of global-local shrinkage priors, which has been shown to scale well to high-dimensional linear models even for sparse training data \citep{bhadra_default_2016, aguilar_R2D2M2_2022}. In particular, such priors can achieve strong regularization of irrelevant coefficients (shrinkage towards zero, i.e. weak sparsity) while providing only mild regularization of truly non-zero coefficients, even if the corresponding predictor variables are highly correlated \citep{bhadra_default_2016, piironen_sparsity_2017}. Below, we will introduce the R2D2 prior in more detail.

The coefficient of determination $R^2$ expresses the proportion of variance explained by a model (here: PCE) in
relation to the total variance of the response $\sigma^2_{\rm total}$. For Gaussian linear models with residual variance $\sigma^2$, we define
\begin{equation}
\label{R2}
R^2 = 1 - \frac{\sigma^2}{\sigma^2_{\rm total}},
\end{equation}
which can be rewritten as
\begin{equation}
R^2 = \frac{\var(\mu)}{\var(\mu) + \sigma^2},
\end{equation}
with linear predictor $\mu$ as defined in \eqref{linpred}. Since the PCE polynomials are by construction orthonormal, we can rewrite the variance of $\mu$ as 
\begin{equation}
\var(\mu) = \sum_{i=1}^{M} c_i^2.
\end{equation}
Now, we exploit this variance decomposition to impose an intuitive joint prior on the PCE coefficients, as any prior on $R^2$ will imply a joint regularization of all the PCE coefficients $c_i$ (excluding the zero-degree coefficient $c_0$).

\begin{figure}
    \centering
    \includegraphics[width=0.99\textwidth]{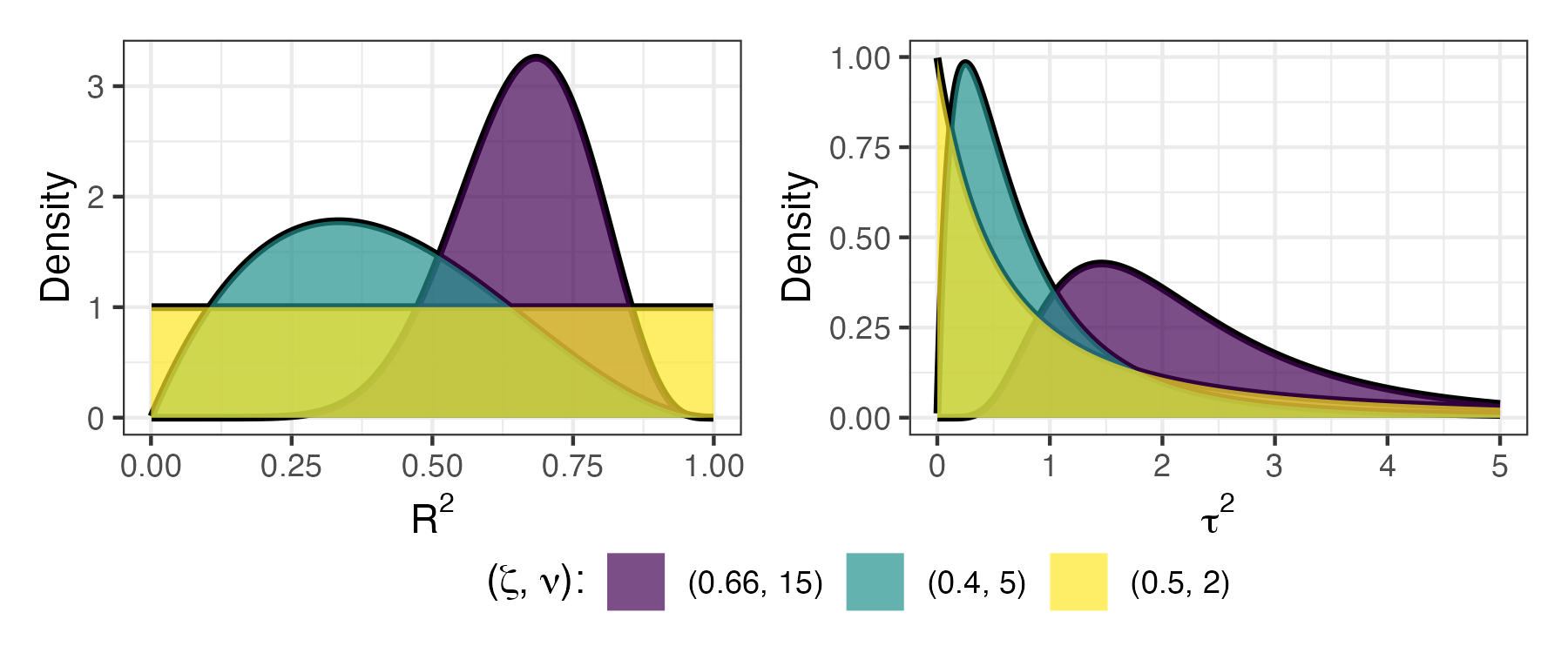}
    \caption{Exemplary densities of the Beta prior for $R^2$ (left) and corresponding BetaPrime prior for $\tau^2 = R^2/(1-R^2)$ (right) with varying mean and precision parameters $(\zeta, \nu)$.}
    \label{fig:beta-priors}
\end{figure}

To set up the R2D2 prior for PCE, we first assign normal priors to the coefficients as $c_i \sim \text{normal}(0, \lambda_i^2)$ and then decompose the prior variance $\lambda_i^2$ of the $i$th coefficient as $\lambda_i^2 = \phi_i \tau^2 \sigma^2$. Here,  $\phi_i$ denotes the proportion of explained variance assigned to the $i$th component, with $\phi = (\phi_1, \ldots, \phi_M)$ forming a simplex, that is, $\phi_{i} \geq 0$ and $\sum_{i=1}^M \phi_{i} = 1$, and $\tau^2 = \frac{R^2}{1-R^2}$ is the variance explained by the PCE. Assigning a Beta prior on $R^2$ with mean parameter $\zeta$ and precision parameter $\nu$ and a Dirichlet prior on $\phi$ with concentration parameter vector $\theta$, we obtain the following complete prior specification \citep{zhang_bayesian_2020}:
\begin{align}
     c_i &\sim \text{normal} \left(0, \phi_i \tau^2 \sigma^2 \right) \quad \forall i = 1, \ldots, M \\
     \tau^2 &= \frac{R^2}{1-R^2} \\
     R^2 &\sim \text{Beta}(\zeta, \nu) \\
     \mathbf{\phi} &\sim \text{Dirichlet}(\mathbf{\theta}) \\
     c_0 &\sim p(c_0) \\
     \sigma^2 &\sim p(\sigma^2)
\end{align}
Different choices of $\zeta$ and $\nu$ imply different a priori assumptions on $R^2$ and subsequently on $\tau$ (which is BetaPrime distributed)  as illustrated in Figure~\ref{fig:beta-priors}. Choosing a uniform prior on $R^2$ via $(\zeta, \nu) = (0.5, 2)$ usually implies sufficient global regularization. However, if the signal to noise ratio is very large (e.g., because training data was generated without noise), larger prior $R^2$ values (as realized via larger $\zeta$) may be favourable as illustrated in Section \ref{CO2-case-study}.

The Dirichlet concentration vector $\mathbf{\theta}$ controls the sparsity of the solution where smaller concentration values imply stronger sparsity \citep{zhang_bayesian_2020, aguilar_R2D2M2_2022}. Here, we set $\mathbf{\theta} = (1, \ldots, 1)$ implying a uniform prior over all simplexes of the corresponding dimension. Smaller values (e.g., $\mathbf{\theta} = (1/2, \ldots, 1/2)$) can be used as well if stronger sparsity is desired, but since we enforce exact sparsity anyway in a second step (see below), we did not further explore this option. Because the prior on $R^2$ controls the global regularization while the prior on $\mathbf{\theta}$ controls the local regularization (sparsity), both priors can be chosen independently of each other. For completeness, densities and a few mathematical properties of the Beta and Dirichlet distributions are provided in Appendix A.

The priors on the zero-degree coefficient $c_0$ and on the residual variance $\sigma^2$ can be assigned separately. In our case studies, we will specify $p(c_0)$ as a normal prior with mean and standard deviation equal to the empirical mean and standard deviation of the training responses, respectively. This represents a weakly-informative prior also chosen as default in the R package brms \cite{brms1}. The prior $p(\sigma^2)$ can be of any common distribution for continuous, non-negative random variables (e.g., an inverse-gamma, half-normal, or half-Cauchy prior). In our case studies, we use a half-normal prior on $\sigma$, with scale parameter appropriately chosen to take both the scale of observed responses and the prior on $R^2$ into account, since $\sigma^2 = \sigma^2_{\rm total} (1 - R^2)$.

\subsection{Statistical methods for variable selection}
\label{variable-selection}

Imposing the R2D2 prior on the PCE coefficients implies a PCE solution that is only weakly sparse in the sense that coefficients identified as irrelevant by the Bayesian PCE model will only be shrunken to values very close to zero, not exactly to zero. Thus, the R2D2 prior alone (or other continuous shrinkage priors) do not result in exactly sparse solutions. Instead, to achieve the latter, variable selection methods are required \citep{piironen_sparsity_2017, piironen_projective_2020}. In general, exact sparsity may be favourable for two primary reasons: (1) If obtaining all the predicting variables comes with large costs of some kind, for example, monetary, ethical, computational, or time-wise; and (2) if the full reference model is not identified or yields bad predictions due to overfitting \citep{piironen_projective_2020}. As evaluation of the PCE polynomials for given input parameters is straightforward, the first motivation does not really apply to PCE -- unless when space or time coordinates at very fine resolutions result in a parallel processing of many coordinate-wise PCE surrogates. What is more, if the prior-induced shrinkage of Bayesian PCE was just right, that is, strong shrinkage was applied to all irrelevant coefficients and comparably little shrinkage to all relevant coefficients, the second motivation would not apply either. However, as the complexity of the problem grows, either by decreasing the number $T$ of training points or by increasing the number $M$ of considered polynomials, the prior-implied shrinkage may become far from optimal, thus either shrinking too much or too little \citep{piironen_sparsity_2017, zhang_bayesian_2020}. In this case, applying a variable selection method in a second step after initially obtaining the full, weakly sparse PCE solution (up to a given degree $d$) may be beneficial to further improve approximation performance of the PCE.

Given any PCE solution, a natural variable selection method is to select the $M_{\rm sel}$ out of all $M$ polynomials with the highest expected (posterior averaged) Sobol indices \eqref{sobol-index}. This is a greedy selection approach in the sense that is does not account for potential dependencies between the polynomials. By construction of the PCE polynomials, they are mutually independent in the limiting case of infinite training data (see Section~\ref{general-PCE}). However, in the practical case of only finite (potentially very small) training data, the polynomials may not be empirically independent for two reasons. First, small dependencies may occur because the chosen finite training points do not perfectly resemble their limiting distribution. Second, and perhaps more severe, for $T$ training points only a maximum $T-1$ polynomials can ever be linearly independent, and so any number of $M \geq T$ polynomials will be necessarily dependent. As we are primarily interested in sparse data scenarios where $M \gg T$, this can be a serious limitation for any greedy variable selection method, as we will also demonstrate in our case studies.

As a fully Bayesian non-greedy variable selection method, we consider the projective prediction (projpred) approach \citep{piironen_projective_2020}, which has been shown to provide superior results compared to alternative procedures \citep{piironen_projective_2020, catalina_projection_2022, catalina_latent_2021, pavone_reference_2020}. In short, projpred aims to find a minimal subset of variables with comparable or better predictive performance than the full reference model (i.e., the PCE solution based on all considered $M$ polynomials) while taking into account the epistemic uncertainty encoded in the posterior distribution of the full model. It does so by aiming to minimize the KL-divergence between the reference model's posterior predictive distribution and the predictive distribution of the sparse sub-model; a problem whose solution is usually intractable. However, for exponential-family models (including PCE expressed as a linear regression model), the KL-minimizing solution can be approximated via maximum-likelihood estimation. This estimation uses the samples from the posterior predictive distribution of the reference model as responses, instead of the actually observed responses \citep{piironen_projective_2020}. Within the projpred framework, a lot of choices can be made by the user, for example, about how to cluster posterior predictive samples of the reference model, how to define the search space of relevant sub-models to investigate, and how to decide when the minimally sufficient subset of variables has been reached \citep{piironen_projective_2020}. 

Here, we consider only a single-cluster solution (all posterior samples projected at once) in combination with a Lasso-type $\ell^1$-regularized model to construct the variable search path \citep{tran_predictive_2012}. Compared to other alternatives in projpred, this is a computationally less demanding option that still yields robust variable selection, although the number of selected variables may be slightly higher than minimally sufficient \citep{piironen_projective_2020}. 
For comparison with greedy Sobol index search and to further reduce computational cost, we fix the number $M_{\rm sel}$ of select polynomials in the sub-models a priori instead of applying more principled but also more expensive decision rules, for example, based on cross-validation. Since we can again regularize the chosen sparse sub-models with an R2D2 prior, selecting a few more terms than minimally sufficient does not do much harm as long as we also select all the relevant polynomials.

\subsection{Fully Bayesian sparse aPC with R2D2 priors}
\label{sparse-bayesian-PCE}

Combining the methods introduced above, we propose a fully Bayesian sparse PCE. It uses the aPC for constructing orthonormal polynomials (see Section~\ref{deterministic-PCE}) and the R2D2 prior to provide joint shrinkage to the PCE coefficients (see Section~\ref{R2D2-prior}). Bayesian sparse PCE models are specified with the R package brms \citep{brms1, brms2} that internally writes highly optimized code in the probabilistic programming language Stan \citep{carpenter2017, stan_2022}. Bayesian estimation of the posterior distribution is conducted by MCMC (see Section~\ref{bayesian-PCE}), more precisely, by the adaptive, recursive Hamiltonian-Monte-Carlo (HMC) algorithm implemented in Stan \citep{stan_2022}. This flavour of MCMC exploits the fact that the (Bayesian) PCE and its posterior distribution in Equation (\ref{posterior}) is differentiable with respect to all PCE parameters. Therefore, it scales well to high-dimensional posteriors \citep{hoffman2014, betancourt2017}. 

In addition to the full PCE models that include all the PCE coefficients up to a given maximal degree, we obtain sparse sub-models either using greedy selection via ordering of the Sobol coefficients or using the projpred method (see Section~\ref{variable-selection}).

\section{Experiments and Results}
\label{case-studies}

In the following subsections, we test and discuss our sparse Bayesian PCE model with R2D2 prior and variable selection on a total of four test cases. These are the Ishigami function, the Sobol function, the signum function and a CO$_2$ benchmark case. All these test cases have been used for testing purposes of PCE concepts and other surrogate approximation techniques. The specific reasoning for the selection of each test case (why it is hard and/or useful, and why it serves specifically to test a sparse form of PCE) is provided in the respective subsections. All code and data required to reproduce the case studies can be found on GitHub (\url{https://github.com/paul-buerkner/Bayesian-sparse-PCE}).

Throughout the case studies, we consider the following three error metrics to evaluate the PCE approximation's performance:
\begin{itemize}
    \item Error of the PCE \emph{mean estimate}
    $\mu(\hat{y}) - \mu(y)$,
    where $\mu(y)$ is the true mean of the target model $\mathcal{M}$ under the given true input measure $\Gamma_{\omega}$ and $\mu(\hat{y})$ is the corresponding PCE approximation. In some cases, we consider variations of this metric, for example, only showing the absolute error  $|\mu(\hat{y}) - \mu(y)|$.
    \item Error of the PCE \emph{standard deviation (SD) estimate}
    $\sigma(\hat{y}) - \sigma(y)$,
    where $\sigma(y)$ is the true SD of the target model $\mathcal{M}$ the given true input measure $\Gamma_{\omega}$ and $\sigma(\hat{y})$ is the corresponding PCE approximation. Like for the mean, we sometimes consider variations of this metric.
    \item The \emph{out-of-sample RMSE} across a sufficiently large number $T_{\rm test}$ of randomly sampled test points (new $\omega$ sampled from the true input measure $\Gamma_{\omega}$) not previously seen by the PCE:
\begin{equation}
\label{RMSE}
    \text{RMSE} = \frac{1}{S} \sum_{s=1}^S \sqrt{ \frac{1}{T_{\rm test}} \sum_{k=1}^{T_{\rm test}} \left( y_{k}-\hat{y}_{k}^{(s)} \right)^2},
\end{equation}
where $y_k$ is the $k$-th response in the test set and $\hat{y}_{k}^{(s)}$ is the $s$th posterior sample of the corresponding PCE approximation out of a total of $S$ posterior samples.

\item Additionally, as MCMC sampling can be time-intensive, we also report the Bayesian PCE estimation times (in minutes).
\end{itemize}

\subsection{Ishigami function}
\label{ishigami-case-study}

As first benchmark, we consider the Ishigami function \citep{ishigami1990}:
\begin{equation}
    y = \sin(\omega_1) + a \sin(\omega_2)^2 + b \, \omega_3^4 \, \sin(\omega_1),
\end{equation}
with $N = 3$ input variables $\omega_1, \omega_2, \omega_3$ being independently uniformly distributed in $[-\pi, \pi]$, and with control parameters $a, b$ being set to $a = 7$ and $b = 0.1$. In this configuration, the Ishigami function has already been applied to benchmark other PCE approaches  \citep{sudret2008}. It serves as a good initial benchmark because its input dimensionality allows for illustrative visualization and it can be nicely approximated by PCE due to the function's smoothness properties. Also, it is a sparse function in the sense that $\omega_2$ has no cross-terms with $\omega_1$ or $\omega_3$, which produces a testable sparsity pattern for our method. What is more, statistics such as mean and variance are analytically known \citep{ishigami1990}.

For training, we consider training data sets with  $T=(10, 25, 50, 100, 200, 286, 400, 800)$ training points. The training points are constructed as the first $T$ values of the 3-dimensional Sobol Sequence. We choose the Sobol sequence as it provides samples more evenly distributed in multidimensional hypercubes than random uniform samples \citep{sobol1967}.

On these training points, we evaluate the Ishigami function as target function for approximation. For simplicity, we linearly scale the prior, the Ishigami function, and the training points to the standard scaling of the Legendre polynomials (i.e., $[-1, 1]$). As PCE basis, we consider all 3-dimensional Legendre polynomials up to a degree of $d = 10$, which implies $M = 286$ polynomials in total. 
These values of $T$ and $M$ were chosen to create a varying range of inference complexity, ranging from strongly under-determined scenarios with $M \gg T$, exactly determined scenarios with $M=T$, to relatively simple over-determined scenarios with $M < T$. We also perform additional experiments with noisy training data, which can be found in Appendix B.

As full reference PCE representation for comparison with the later-computed sparse PCEs, we used all $M$ polynomials as PCE basis terms. We equip the corresponding PCE coefficients with an R2D2 prior on the $M - 1$ non-constant polynomials as described in Section \ref{sparse-bayesian-PCE}. Specifically, we used an R2D2 prior with $R^2 \sim \text{Beta}(\zeta = 0.5, \nu = 2)$ that implies a-priori uniformity of $R^2$ in $[0, 1]$ and provides some overall regularization while still allowing high $R^2$ to be possible.

Based on this full PCE Ansatz, we construct one PCE representation for each training set of size $T$. Then, for each full PCE representation, we fitted two sparse sub-PCEs each that merely differ in sparse selection method. The first uses the ordered Sobol indices \citep{sudret2008} and the second uses projective prediction (projpred) for variable selection \citep{piironen_projective_2020}. For each selection method, we extracted the $M_{\rm sel} = 25$ most important polynomial terms out of all non-constant ones plus the constant, $0th$-degree polynomial. Only the selected polynomials were included in the respective sparse PCEs. For these selected polynomials and their corresponding coefficients, we again used a uniform R2D2 prior with $R^2 \sim \text{Beta}(\zeta = 0.5, \nu = 2)$. This still provides some regularization on the PCE coefficients, even for the already sparse models. Even though this is the same prior on $R^2$ as for the reference models, the implied regularization on each of the sparse models' coefficients is comparably milder because the explained variance is distributed among much fewer coefficients. 

\begin{figure}
    \centering
    \includegraphics[width=0.99\textwidth]{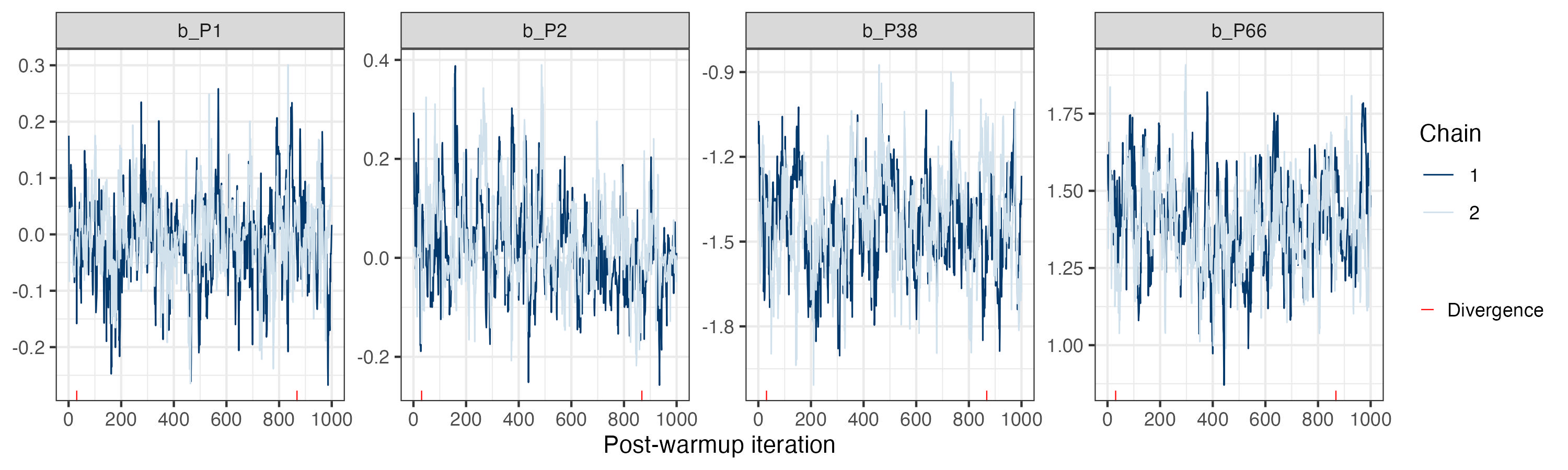}
   \caption{Trace plot of four selected regression coefficients (two irrelevant and two relevant) for the full reference model on the Ishigami function. This shows good convergence of the reference model. The y-axis depicts the corresponding coefficient values.}
   \label{fig:ishigami-trace}
\end{figure}

To fit these PCE models to the training data sets, we produced posterior samples using MCMC as explained in the previous section. 
Two independent Markov Chains were run each for 2000 iterations. From these, the first 1000 were discarded as warmup, leading to a total of 2000 post-warmups samples used for inference. All models converged well as indicated by standard algorithm-specific \citep{betancourt2017} and algorithm-agnostic \citep{vehtari2021} convergence diagnostics. Trace plots of the PCE coefficients illustrate excellent mixing of the MCMC chains also graphically (see Figure \ref{fig:ishigami-trace} for trace plots of four selected PCE coefficients of the reference model based on $T=100$ training points).

\begin{figure}
    \centering
    \includegraphics[width=0.8\textwidth]{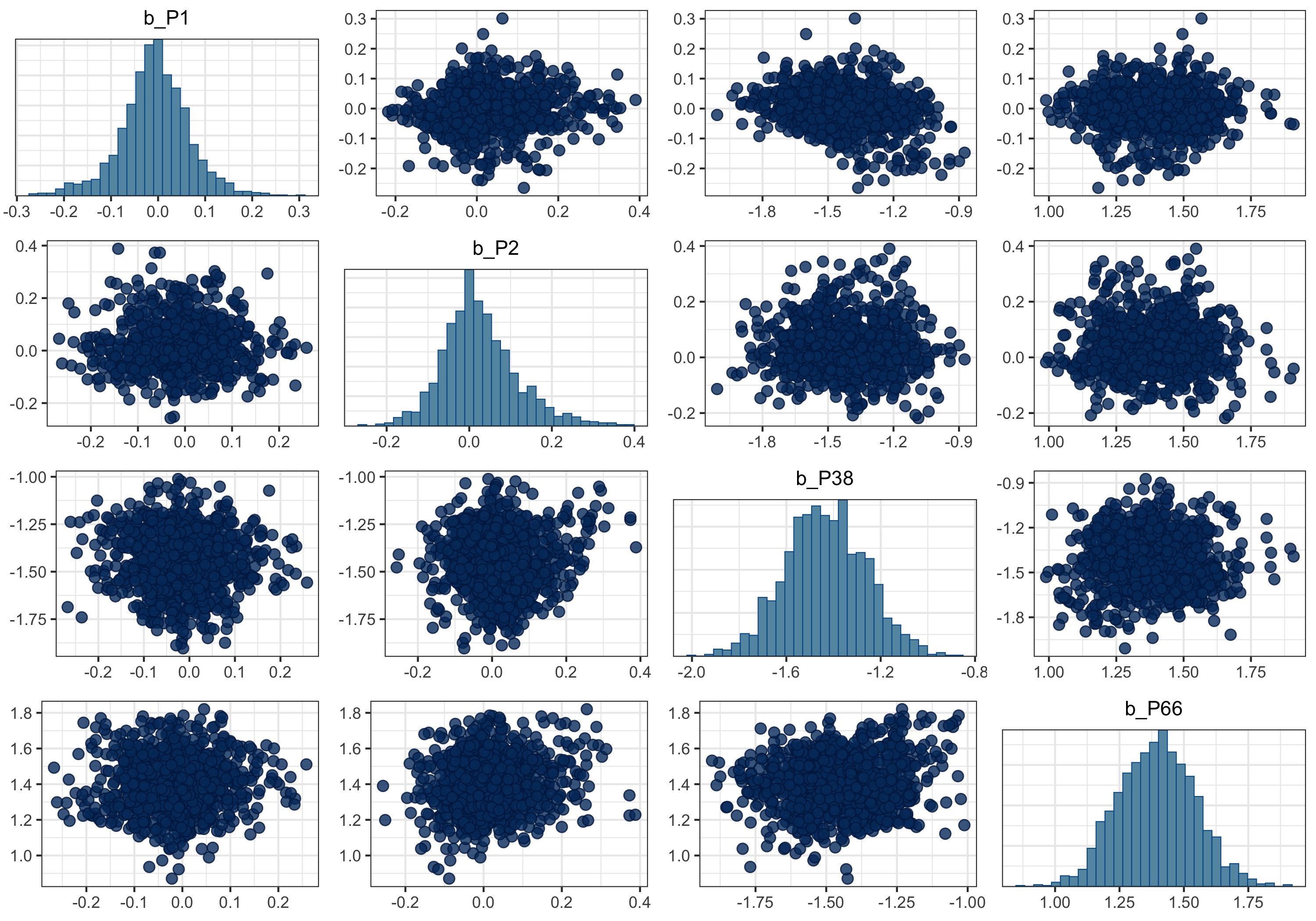}
    \caption{Pairs plot of four selected regression coefficients (two irrelevant and two relevant) for the full reference model on the Ishigami function based on $T=100$ training points. This illustrates (a) approximate pairwise independence of the coefficients even for $M > T$ and (b) strong shrinkage for irrelevant coefficients. In the diagonal plots, the x-axis depicts the coefficient values and the y-axis their binned probabilities. In the off-diagonal plots, the axes depict coefficients values of the corresponding column and row, respectively.}
    \label{fig:ishigami-pairs}
\end{figure}

Posterior histograms of the reference model's PCE coefficients demonstrate the characteristic shrinkage properties of the R2D2 prior as the posteriors of irrelevant coefficients are strongly centered around zero with heavier-than-Gaussian tails (see the first two diagonal elements of Figure \ref{fig:ishigami-pairs}). What is more, the bivariate pairs plots in Figure \ref{fig:ishigami-pairs} show approximate pairwise independence of the PCE coefficients' posteriors. This is caused by the independence of the corresponding Legendre polynomials.

\begin{figure}
    \centering
    \includegraphics[width=0.99\textwidth]{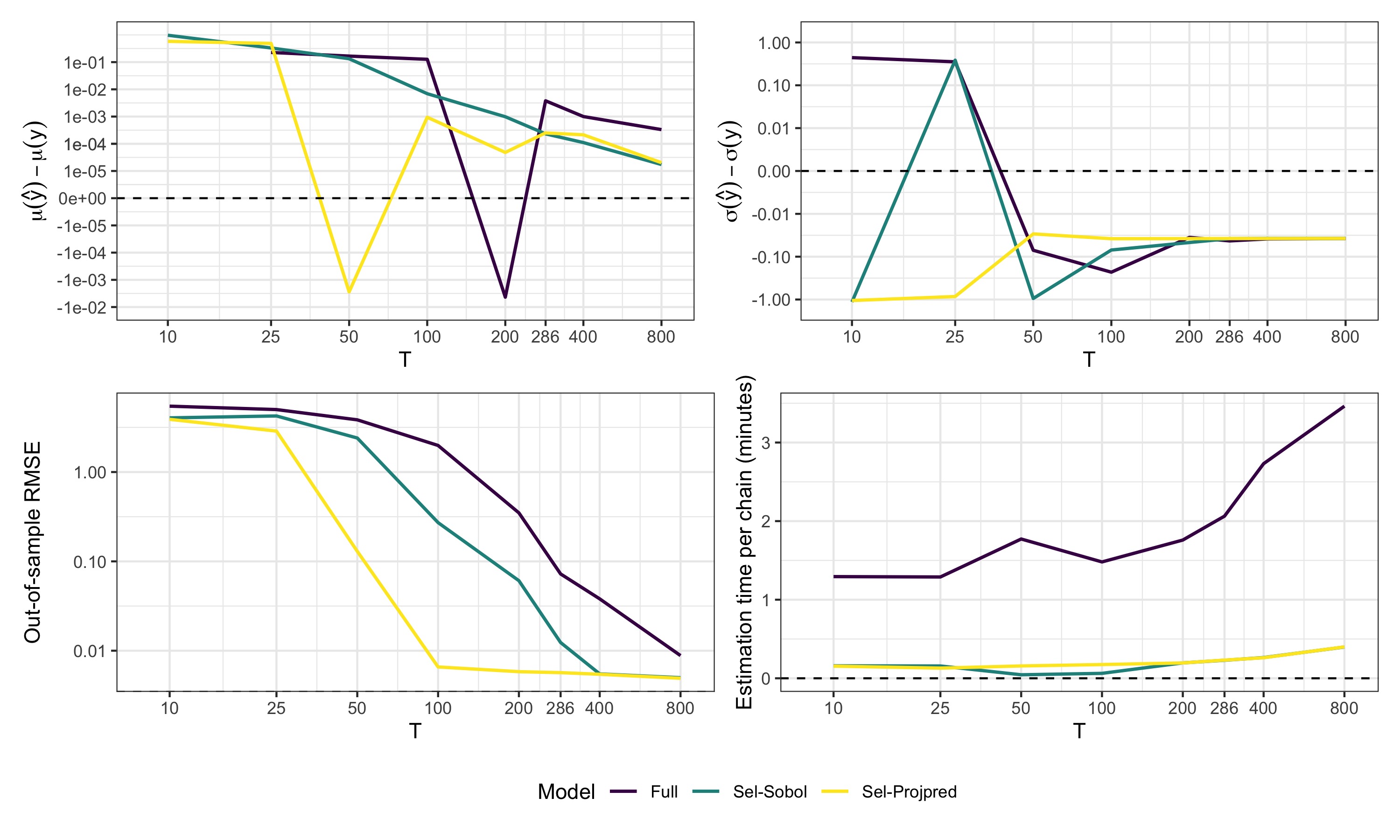}
    \caption{Summarized results for the Ishigami function $(a=7, b=0.1)$ by the number of training points $T$ (log-scaled; evaluated sizes are equal to the displayed axis ticks), and model type (colors). Top-left: bias of the mean $\mu$; top-right: bias of the standard deviation $\sigma$; bottom-left: average out-of-sample predictive RMSE; bottom-right: average estimation times per fitted model.}
    \label{fig:ishigami-summaries}
\end{figure}

From Figure~\ref{fig:ishigami-summaries}, we see that the full reference model is mostly competitive with the sparse sub-models in terms of mean and SD bias. However, it has uniformly worse out-of-sample RMSE especially for small $T$. This demonstrates that the R2D2-implied regularization alone (i.e., before selection of sparse subsets) may be insufficient to achieve satisfactory function approximation in $M \gg T$ scenarios, at least under the conditions considered here. In contrast, both sparse sub-models perform much better. Specifically the sparse projpred model achieves very good out-of-sample RMSE, while only slightly underestimating the true Ishigami SD already for $T=50$ training points (again see Figure~\ref{fig:ishigami-summaries}).

\begin{figure}
    \centering
    \includegraphics[width=0.99\textwidth]{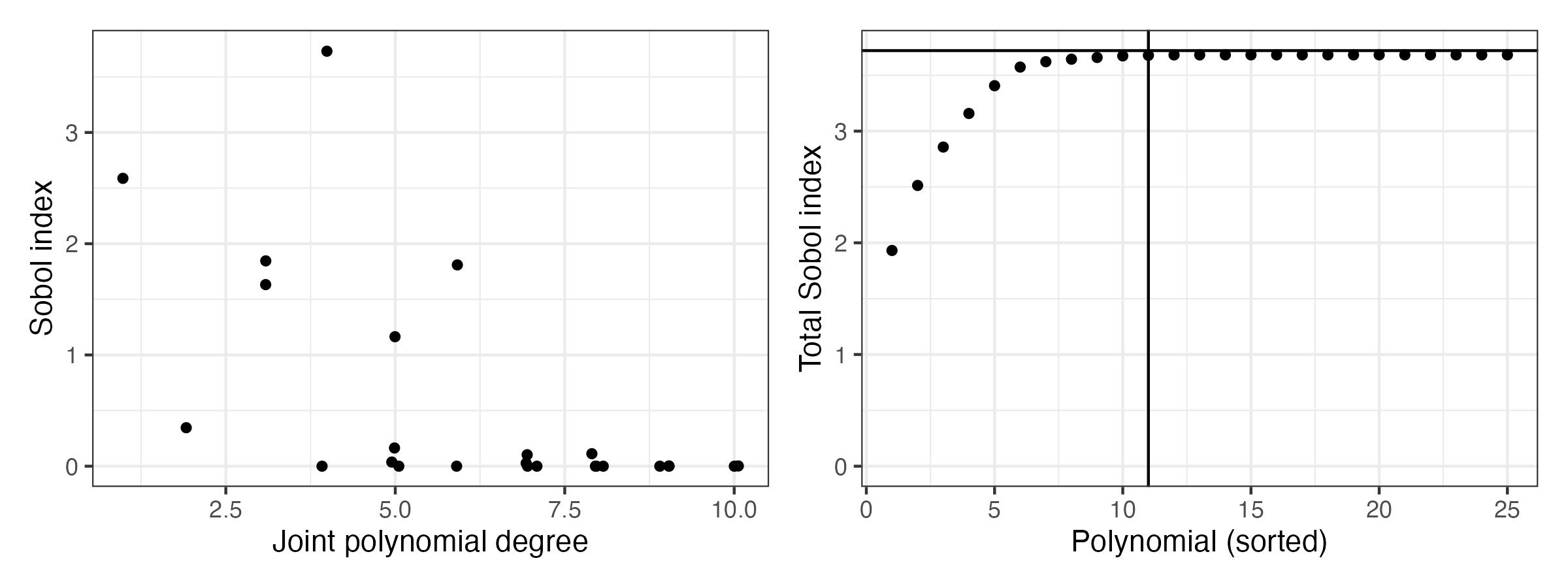}
    \caption{Posterior mean Sobol indices and total Sobol indices for the sparse projpred model on the Ishigami function based on based on $T=100$ training points and the $M_S = 25$ most important non-constant polynomials. The vertical line indicates the first index with only $0.5\%$ or less increase in the total Sobol index. Error bars indicate 99\% posterior credible intervals but they are too small to be clearly visible. }
    \label{fig:ishigami-poly-degree-varsel}
\end{figure}

\begin{figure}
    \centering
    \includegraphics[width=0.99\textwidth]{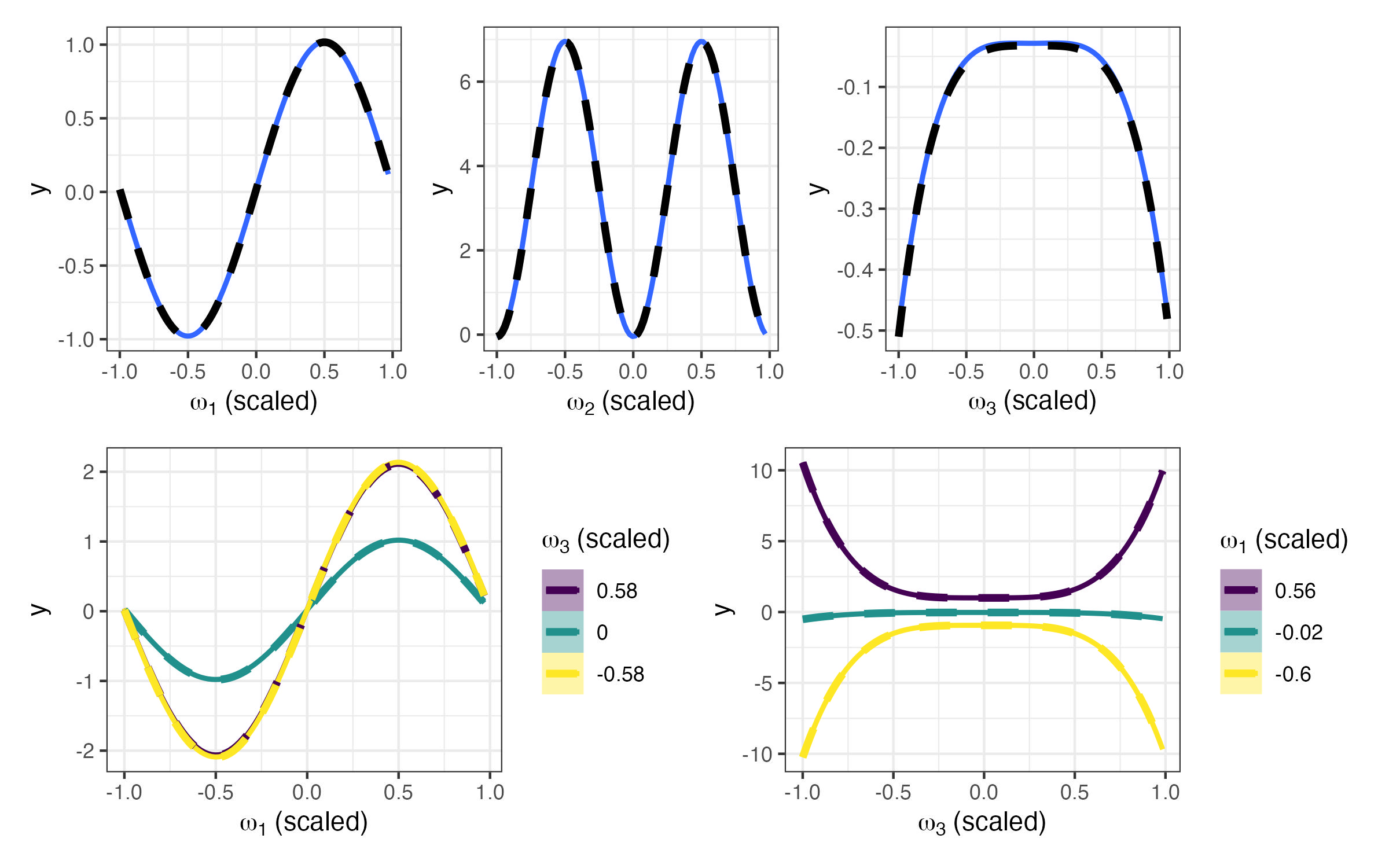}
    \caption{Conditional prediction for the sparse projpred model of the Ishigami function based on $T=100$ training points and the $M_S = 25$ most important polynomials. In each plot, all non-displayed input variables are fixed to their mean value. Dashed lines indicate the corresponding true Ishigami function values. Shaded areas indicate 95\% posterior credible intervals but they are too small to be clearly visible. The conditional predictions match the true values almost perfectly.}
    \label{fig:ishigami-ceffects}
\end{figure}

For the sparse projpred model based on $T=100$ training points, Sobol and total Sobol indices are shown in Figure \ref{fig:ishigami-poly-degree-varsel}. As can be seen, only $10$ non-constant polynomials (plus the constant polynomial) seem to be sufficient to achieve an almost perfect approximation. Table~\ref{tab:ishigami-poly-small-varsel} shows more details on the corresponding polynomial degrees), and one can see that, as ideally expected by knowing the true underlying Ishigami function, no cross-terms with $\omega_2$ occur.
The impression of a good approximation is confirmed in Figure \ref{fig:ishigami-ceffects}, showing conditional predictions of the sparse projpred model against the corresponding true Ishigami function values.
As $T$ increases further, the Ishigami mean and SD estimates do not improve anymore compared to the $T=100$ sparse projpred model, presumably because the remaining approximation error is due to polynomials of degree higher than $d=10$, which were not included in the PCE approximation here.

Together, these results demonstrate that, with our proposed methods, performing global polynomial search for sparse selection of PCE terms (up to a fixed maximal degree) is possible and can lead to good approximations even if $M > 6 T$. What is more, nearly perfect approximations can already be achieved for $T=100$, which corresponds to $M \approx 3 T$ in the scenario considered here.


\begin{table}
\centering
\begin{tabular}[t]{llrrr}
\hline
Mean & 95\%-CI & $d(\omega_1)$ & $d(\omega_2)$ & $d(\omega_3)$ \\ 
\hline
3.73 & [3.73, 3.73] & 0 & 4 & 0\\
2.59 & [2.58, 2.59] & 1 & 0 & 0\\
1.85 & [1.84, 1.85] & 1 & 0 & 2\\
1.81 & [1.81, 1.81] & 0 & 6 & 0\\
1.63 & [1.63, 1.64] & 3 & 0 & 0\\
1.16 & [1.16, 1.17] & 3 & 0 & 2\\
0.35 & [0.34, 0.35] & 0 & 2 & 0\\
0.16 & [0.16, 0.16] & 1 & 0 & 4\\
0.11 & [0.11, 0.11] & 0 & 8 & 0\\
0.10 & [0.10, 0.10] & 3 & 0 & 4\\
\hline
\end{tabular}
\caption{Posterior mean Sobol indices (Mean), 95\% posterior credible intervals (95\%-CI), and polynomial degrees $d(\omega_m)$ by input variable of the 10 most important non-constant polynomials as selected by the sparse projpred model for the Ishigami function based on $T=100$ training points.}
\label{tab:ishigami-poly-small-varsel}
\end{table}

\newpage
\subsection{Sobol function}
\label{sobol-case-study}

As second benchmark, we consider the Sobol function \citep{sobol1967,sobol2003}:
\begin{equation}
\label{sobol-function}
    y = \prod_{j=1}^N \frac{|4 \omega_j - 2| + a_j}{1+a_j},
\end{equation}
with $N = 8$ input variables $\omega_1, \ldots, \omega_8$ being independently uniformly distributed in $[0, 1]$, and with its control parameter vector $a$ being set to $a = (1, 2, 5, 10, 20, 50, 100, 500)$. In this configuration, the Sobol function has already been applied to benchmark other PCE approaches in previous studies \citep[e.g.,][]{sudret2008}. As can be readily seen from Equation~\ref{sobol-function}, this function is symmetric around $\omega_j = 0.5$. Thus, only even-degree polynomials should contribute to its approximation. Accordingly, any sparse PCE method performing iterative local search (i.e., stopping their search as soon as they cannot find a polynomial of a given degree) would fail immediately: because there are no odd-degree polynomials contributing to the approximation, such a search would stop at $0th$, $2nd$, $4th,...$ degree, depending on the initialization. Additionally, the Sobol function is an interesting benchmark because it has a relatively high-dimensional input space ($N=8$ in our case) and is not continuously differentiable in $\omega_j = 0.5$, thus making it harder to approximate this function via PCE.

Similar to the Ishigami test case, we provide $T=(100, 300, 900, 1001, 2700, 3003, 8100)$ training points given by the first $N$ values of the 8-dimensional Sobol Sequence \citep{sobol1967} for training the upcoming PCE approximations. On these training points, we evaluated the Sobol function to provide actual training data.

As a first non-sparse PCE basis, we consider all 8-dimensional Legendre polynomials (scaled to fit the input range) up to a total degree of $d = 6$. This leads to $M = 3003$ possible polynomials in total. Additionally, we use a second set of polynomials as PCE basis, this time only considering the first $4$ input variables in the PCE and corresponding 4-dimensional Sobol Sequence training points. In contrast to the first version, we now consider all 4-dimensional Legendre polynomials up to a total degree of $d = 10$, which implies $M = 1001$ polynomials in total. The underlying true Sobol function remains $8$-dimensional, i.e. the training outcomes stay the same. This means that the last $4$ input parameters are still random variables, but that the PCE assumes them to be irrelevant/constant. This is a valid choice, because the last $4$ input variables are comparably less important in the Sobol function, and instead the basis approximates the relevant variables up to degree $d=10$ instead of only $d=6$. These two PCE bases explore the trade-off between the number of considered input variables and the maximal degree of polynomials included in the approximation. 

With both PCE bases and the different sizes $N$ of training data, we again create a varying range of inference complexity, as we already did for the Ishigami test case. Arguably, our second set of PCE models $(M=1001)$ constitutes an overall simpler scenario that the first set $(M=3003)$, because it has only about $1/3$ of the number of polynomials in total.

Prior choices, sparsifying procedures, and MCMC sampling methods were the same as for the Ishigami case study (see Section~\ref{ishigami-case-study}). Again, all models converged well as indicated by standard algorithm-specific and algorithm-agnostic convergence diagnostics.

\begin{figure}
    \centering
    \includegraphics[width=0.99\textwidth]{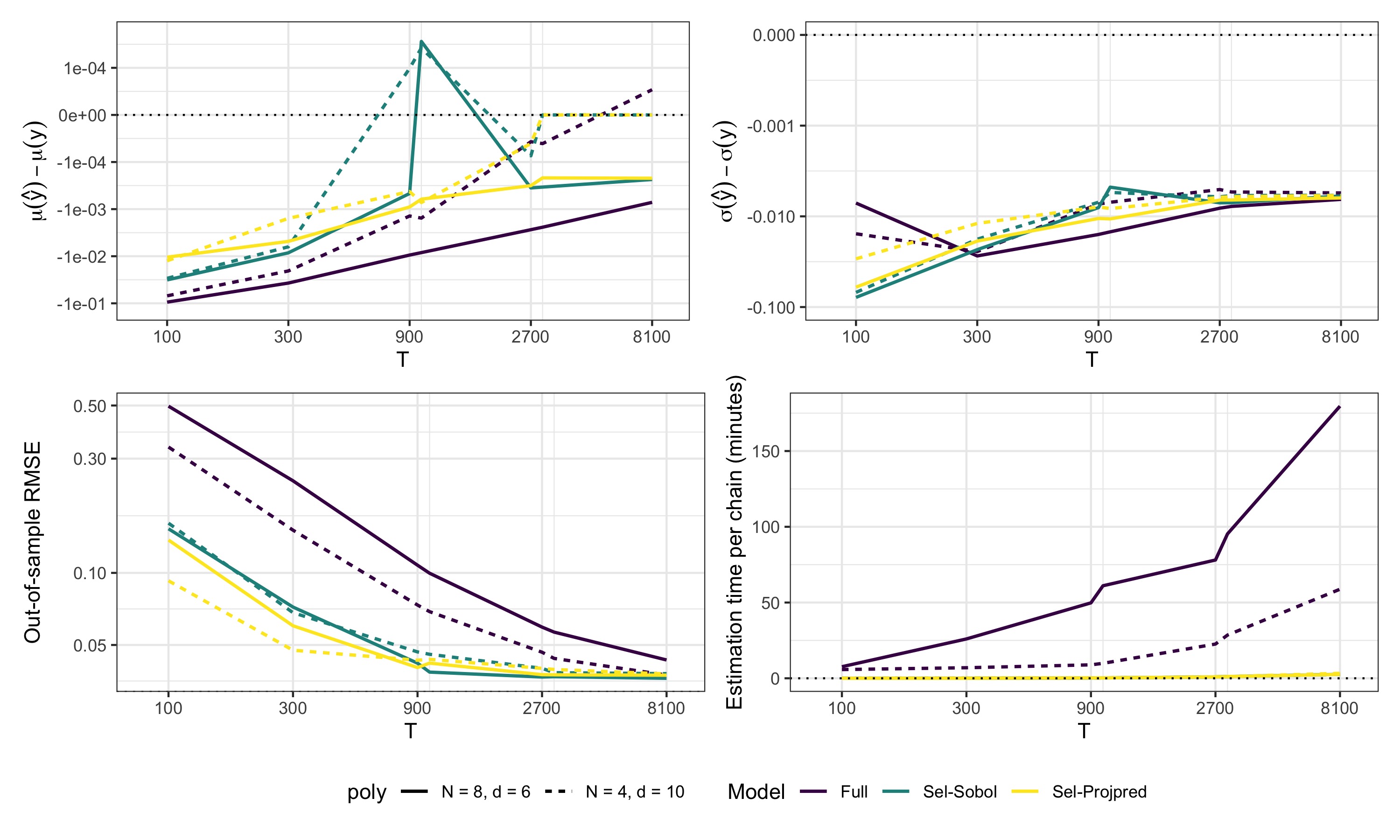}
    \caption{Summarized results for the Sobol function ($a=(1, 2, 5, 10, 20, 50, 100, 500)$) by the number of training points $T$ (log-scaled; evaluated sizes are equal to the displayed axis ticks), model type (colors), and considered polynomials (line type). Top-left: bias of the mean $\mu$; top-right: bias of the standard deviation $\sigma$; bottom-left: average out-of-sample predictive RMSE; bottom-right: estimation times per fitted model.}
    \label{fig:sobol-summaries}
\end{figure}

The approximation performance of the resulting PCEs was evaluated by comparing their approximations of the Sobol function's mean and SD to the corresponding true values \citep{sobol1967, sobol2003}. Additionally, we computed the out-of-sample RMSE across $T_{\rm test} = 500$ randomly sampled test points not previously seen by the models as per Equation~\eqref{RMSE}. From Figure~\ref{fig:sobol-summaries} we see that, similar to the Ishigami case, the full reference model is mostly competitive with the sparse sub-models in terms of mean and SD bias. But, again, it has worse out-of-sample RMSE, especially for small $T$. This demonstrates again that the R2D2-implied regularization alone (i.e., before sparse selection) may be insufficient to achieve satisfactory function approximation in $M \gg T$ scenarios. In contrast, both sparse sub-models perform much better, with the projpred models having a slight edge for smaller $N$ (again see Figure~\ref{fig:ishigami-summaries}). As compared to the models considering all 8 input parameters, the models using only the first 4 variables show better mean approximation overall, better SD approximation and RMSE, as well as shorter estimation times for the full reference model.

\begin{figure}
    \centering
    \includegraphics[width=0.99\textwidth]{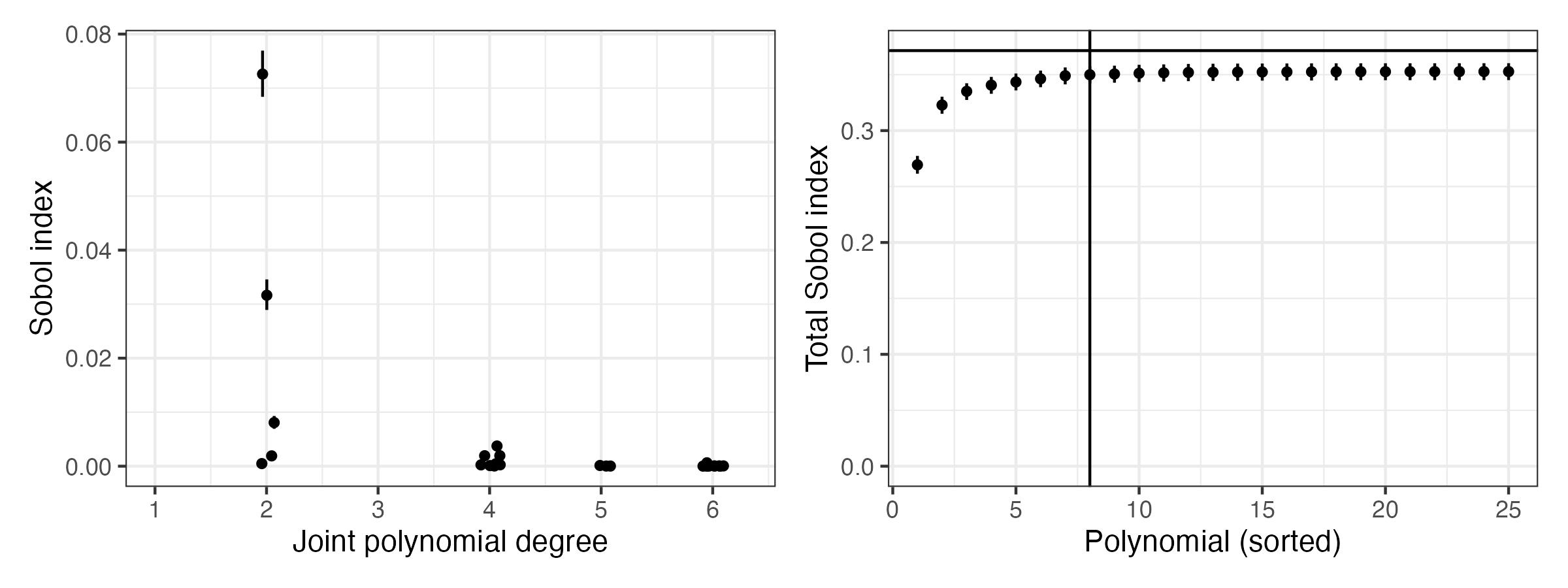}
    \caption{Posterior mean Sobol indices and total Sobol indices for the sparse projpred model ($T=300$) on the Sobol function ($N = 8$ variables) using the $M_S = 25$ most important non-constant polynomials constructed from all 8 variables. The vertical line indicates the first index with only $0.5\%$ or less increase in the total Sobol indices. Error bars indicate 99\% CIs.}
    \label{fig:sobol-poly-degree-M8-varsel}
\end{figure}

\begin{figure}
    \centering
    \includegraphics[width=0.99\textwidth]{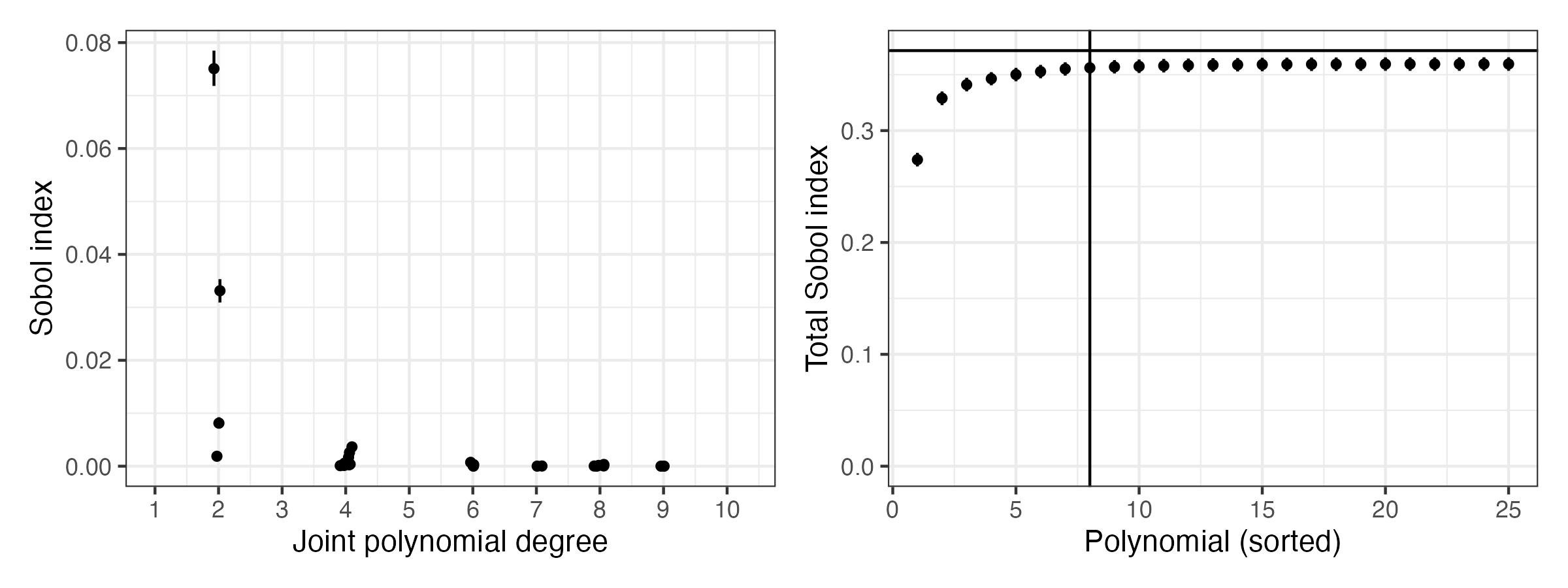}
    \caption{Posterior mean Sobol indices and total Sobol indices for the sparse projpred model ($T=300$) on the Sobol function ($N = 8$ variables) using the $M_S = 25$ most important non-constant polynomials constructed from the first 4 variables only. The vertical line indicates the first index with only $0.5\%$ or less increase in the cumulative Sobol indices. Error bars indicate 99\% CIs but they are too small to be clearly visible in most cases.}
    \label{fig:sobol-poly-degree-M4-varsel}
\end{figure}

For the sparse projpred models based on $T=300$ training points, Sobol and cumulative Sobol indices are shown in Figures  \ref{fig:sobol-poly-degree-M8-varsel} and \ref{fig:sobol-poly-degree-M4-varsel} using all 8 or only the first 4 input parameters, respectively. As can be seen from the Figures, only $8$ polynomials (including the constant polynomial) seem to be sufficient to achieve a good approximation of the Sobol function; and more polynomials (within the considered degrees) do not further improve this approximation. All of the relevant polynomials are even, as one would expect because of the symmetry of the Sobol function in terms of total Sobol indices (see Tables \ref{tab:sobol-poly-small-M8-varsel} and \ref{tab:sobol-poly-small-M4-varsel} for details on the polynomial degrees).

Within the total of $M_{\rm sel} = 25$ selected polynomials, only very few are odd. These few odd terms are included simply because $M_{\rm sel}$ is larger than the number of actually relevant polynomials. As can be seen in Figures  \ref{fig:sobol-poly-degree-M8-varsel} and \ref{fig:sobol-poly-degree-M4-varsel}, those odd polynomials are correctly shrunken to basically zero with almost no remaining uncertainty. 
Going back to Figure~\ref{fig:ishigami-summaries}, as $T$ increases, the PCE estimates for the mean and SD of the Sobol function continue to improve slowly as compared to the sparse $T=300$ models. However, the SD bias flattens out at around $\text{bias}(\sigma) = -0.1$. This asymptotic performance is about the same for both sets of models. So considering all 8 input parameters, but ignoring polynomials above degree $d=6$ leads to about the same asymptotic SD bias as considering only the first 4 variables but ignoring all polynomials above degree $d=10$. As can be seen from Tables \ref{tab:sobol-poly-small-M8-varsel} and \ref{tab:sobol-poly-small-M4-varsel}, this is because both of these models identify the same polynomials as relevant, all of which are constructed from the first 4 input variables and have degree 6 or less.

Together, these results demonstrate that with our proposed methods, performing global polynomial search (up to a fixed maximal degree) is possible and can lead to good approximations even if $M > 30 T$. In specific, we showed that an "even-degree only" function is correctly identified -- an achievement impossible for local-type greedy search scheme in sparse PCE methods. What is more, very close approximations can already be achieved for $T=300$, which corresponds to $M \approx 10 T$ in the models considering all $N=8$ input variables. From the comparsion between $N=8$ at moderate polynomial degree ($d=6)$ and $N=4$ at higher polynomial degree ($d=10)$, we learn that some form of manual trial-and-error for an appropriate search space can be helpful. The reason is that the global search is highly beneficial, but the corresponding search space still grows combinatorically fast with $N$ and $d$.


\begin{table}
\centering
\begin{tabular}[t]{llrrrrrrrr}
\hline
Estimate & 95\%-CI & $d(\omega_1)$ & $d(\omega_2)$ & $d(\omega_3)$ & $d(\omega_4)$ & $d(\omega_5)$ & $d(\omega_6)$ & $d(\omega_7)$ & $d(\omega_8)$ \\
\hline
0.073 & [0.069, 0.076] & 2 & 0 & 0 & 0 & 0 & 0 & 0 & 0\\
0.032 & [0.029, 0.034] & 0 & 2 & 0 & 0 & 0 & 0 & 0 & 0\\
0.008 & [0.007, 0.009] & 0 & 0 & 2 & 0 & 0 & 0 & 0 & 0\\
0.004 & [0.003, 0.004] & 4 & 0 & 0 & 0 & 0 & 0 & 0 & 0\\
0.002 & [0.001, 0.002] & 2 & 2 & 0 & 0 & 0 & 0 & 0 & 0\\
0.002 & [0.001, 0.002] & 0 & 4 & 0 & 0 & 0 & 0 & 0 & 0\\
0.002 & [0.001, 0.002] & 0 & 0 & 0 & 2 & 0 & 0 & 0 & 0\\
\hline
0.001 & [0.000, 0.001] & 6 & 0 & 0 & 0 & 0 & 0 & 0 & 0\\
0.000 & [0.000, 0.001] & 0 & 0 & 0 & 0 & 2 & 0 & 0 & 0\\
0.000 & [0.000, 0.001] & 2 & 0 & 2 & 0 & 0 & 0 & 0 & 0\\
0.000 & [0.000, 0.000] & 0 & 0 & 4 & 0 & 0 & 0 & 0 & 0\\
0.000 & [0.000, 0.000] & 0 & 2 & 2 & 0 & 0 & 0 & 0 & 0\\
\hline
\end{tabular}
\caption{Posterior mean Sobol indices (Mean), 95\% posterior credible intervals (95\%-CI), and polynomial degree $d(\omega_j)$ by input variable of the 12 most important non-constant polynomials constructed from all 8 input variables as selected by the sparse projpred model for the Sobol function ($N=8$ variables) based on $T=300$ training points. All of these polynomials are even, but beyond the 7 most important of them, none of the further polynomials contributes to the PCE approximation in relevant manner.}
\label{tab:sobol-poly-small-M8-varsel}
\end{table}

\begin{table}
\centering
\begin{tabular}[t]{llrrrr}
\hline
Estimate & 95\%-CI & $d(\omega_1)$ & $d(\omega_2)$ & $d(\omega_3)$ & $d(\omega_4)$ \\
\hline
0.075 & [0.072, 0.078] & 2 & 0 & 0 & 0\\
0.033 & [0.031, 0.035] & 0 & 2 & 0 & 0\\
0.008 & [0.007, 0.009] & 0 & 0 & 2 & 0\\
0.004 & [0.003, 0.004] & 4 & 0 & 0 & 0\\
0.003 & [0.002, 0.003] & 2 & 2 & 0 & 0\\
0.002 & [0.001, 0.002] & 0 & 0 & 0 & 2\\
0.002 & [0.001, 0.002] & 0 & 4 & 0 & 0\\
\hline
0.001 & [0.000, 0.001] & 6 & 0 & 0 & 0\\
0.001 & [0.000, 0.001] & 2 & 0 & 2 & 0\\
0.000 & [0.000, 0.001] & 0 & 0 & 4 & 0\\
0.000 & [0.000, 0.001] & 8 & 0 & 0 & 0\\
0.000 & [0.000, 0.000] & 0 & 6 & 0 & 0\\
\hline
\end{tabular}
\caption{Posterior mean Sobol indices (Mean), 95\% posterior credible intervals (95\%-CI), and polynomial degree $d(\omega_j)$ by input variable of the 12 most important non-constant polynomials constructed from the first $4$ input variables ($\omega_1$ to $\omega_4$) as selected by the sparse projpred model for the Sobol function ($M=8$ variables) based on $T=300$ training points. All of these polynomials are even, but beyond the 7 most important of them, none of the further polynomials contributes to the PCE approximation in relevant manner.}
\label{tab:sobol-poly-small-M4-varsel}
\end{table}



\newpage
\subsection{Signum function}
\label{signum-case-study}
\begin{figure}
    \centering
    \includegraphics[width=0.99\textwidth]{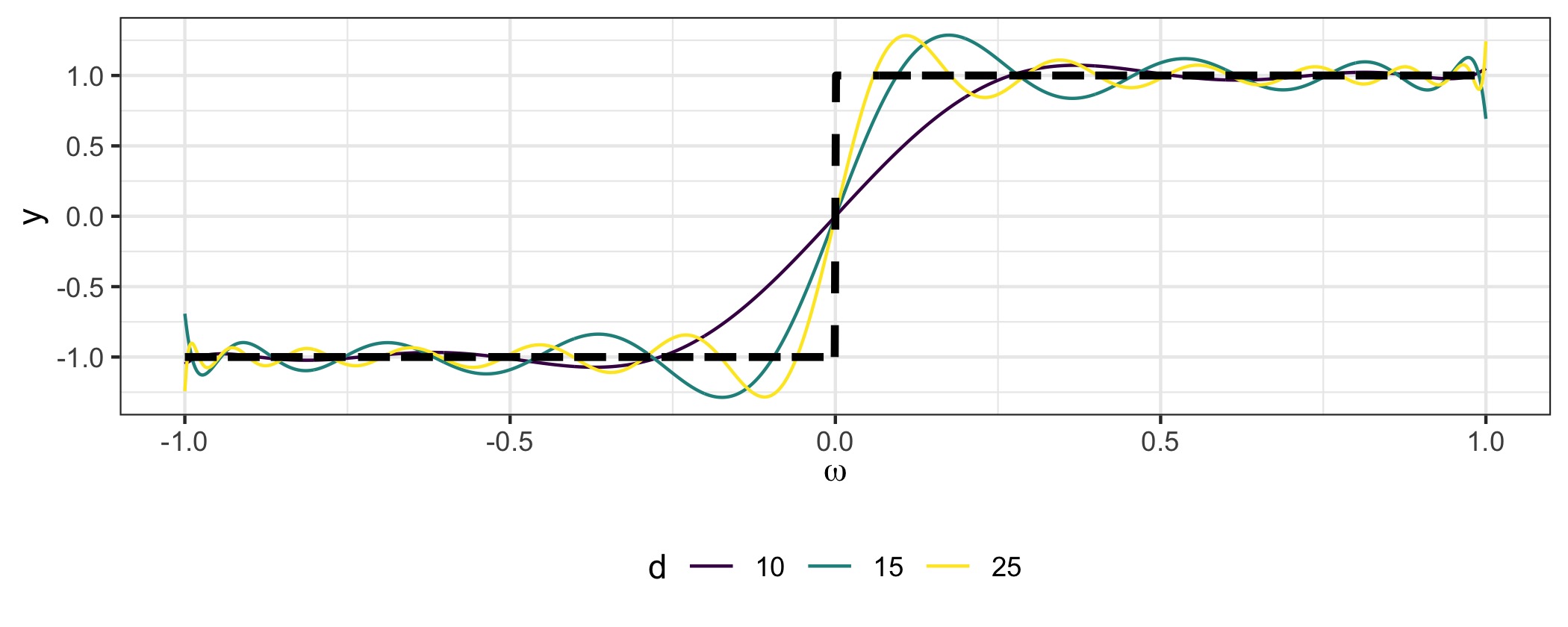}
    \caption{Standard PCE expansion of the Signum function for $d=(10,15,25)$ and $T=d+1$ training points given by Gaussian integration. The dashed line indicates the Signum function.}
    \label{fig:signum-pce}
\end{figure}

Polynomial-based surrogate models such as PCE are highly popular approximators of complex target functions.
If the target functions are continuous,
PCE ensures uniform convergence on compact sets and at least $L^2$-convergence for functions in $L^2$ according to the Weierstrass theorem \cite{MR3364576}.
Nevertheless, these theoretical requirements are not always given in many practical applications.
In particular, two effects are responsible for many drawbacks in the accuracy 
of the polynomial-based expansions:
It is known that insufficient smoothness of the target function as well as selection of sub-optimal evaluation (training) points can lead to oscillations of the response surface, 
the so called Gibbs' and Runge phenomena \cite{MR3364576}.
More precisely, the Gibbs' phenomenon describes oscillations related to the discontinuities, 
and the Runge phenomenon describes oscillations caused by the selection of the evaluation points. 
One of the classical examples of the Gibbs' phenomenon is the approximation of the Signum (i.e., sign) function as displayed in Figure~\ref{fig:signum-pce}. Here, we can see that the increasing polynomial degree increases the number of oscillation and does not necessarily reduce their amplitude. As the signum function is one-dimensional, we can fully visualize the problem. Also, by construction, it is a purely odd function (unlike the Sobol function is purely even), and again provides an easily testable sparsity pattern. 

In this case study, we consider PCEs of polynomial degree $d=(1, \ldots, 10)$
with both $T=M=d+1$ training points and polynomials. There are given either by Gaussian quadrature points to investigate the effect on the Gibbs' phenomenon around the discontinuity at $0$. Or they are given by the Sobol sequence that causes the Runge phenomenon on the right-hand side of the input space (i.e., close to $1$).
We deliberately investigate these $T=M$ cases as those are the minimal-data scenarios where standard PCE expansion is directly solvable (see Section~\ref{deterministic-PCE}). In addition to Standard PCE, we estimated Bayesian sparse PCE with the R2D2 prior (\emph{Bayesian-R2D2}) and Bayesian PCE with improper flat priors on the coefficients (\emph{Bayesian-flat}). The latter differs from Bayesian-R2D2 only in this choice of prior and can also be understood as a sampling-based equivalent of Standard PCE. Thus, the Bayesian-flat approach can be used to disentangle the influence of the R2D2 prior from the noise introduced by the MCMC algorithm. The settings of the MCMC algorithm were the same as in the Ishigami case study (see Section~\ref{ishigami-case-study}). As the purpose of this case study was only to study the regularizing effect of the R2D2 prior in case of a non-smooth target function, no further variable selection was performed.

\begin{figure}
    \centering
    \includegraphics[width=0.99\textwidth]{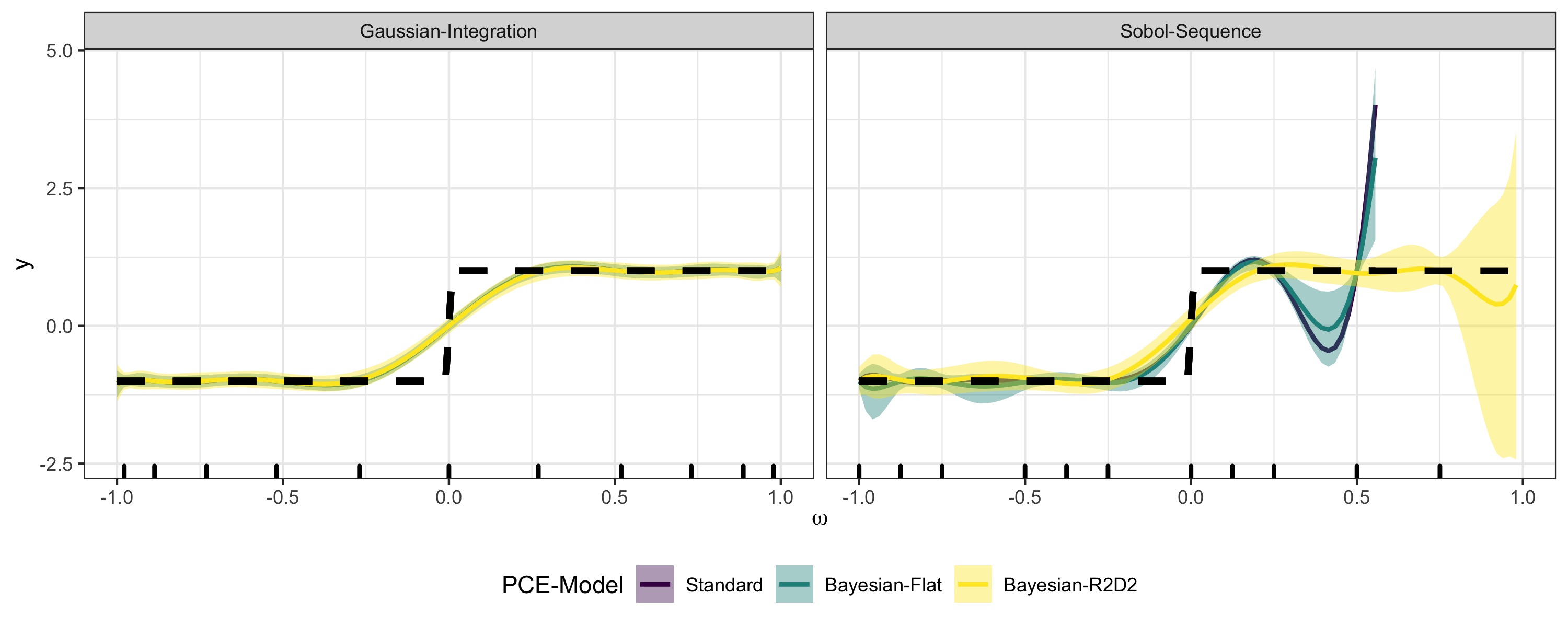}
    \caption{Conditional predictions for different PCE models of the Signum function based on $M=11$ polynomials and $T=11$ training points, either given by Gaussian integration (left) or the Sobol Sequence (right). The y-axis was truncated at $5$ from above and at $-5$ from below to keep the plot readable. No predictions are shown after the first exceedance of these thresholds for a given PCE model. The dashed line indicates the Signum function. The small vertical dashes on the x-axis indicate the training point locations.}
    \label{fig:signum-summaries}
\end{figure}

In Figure~\ref{fig:signum-summaries}, we see exemplary for the case of $d=10$ and  $T=11$ that the surrogate-models provided by
classical PCE, Bayesian-Flat and Bayesian-R2D2 models
trained on Gaussian quadrature points show very similar behaviour, at least visually.
In contrast, on the training points given by the Sobol sequence, the Bayesian-R2D2 models
provide a significantly more reliable surrogate model than the other two here-considered PCE models in areas with sparse training points.

\begin{figure}
    \centering
    \includegraphics[width=0.99\textwidth]{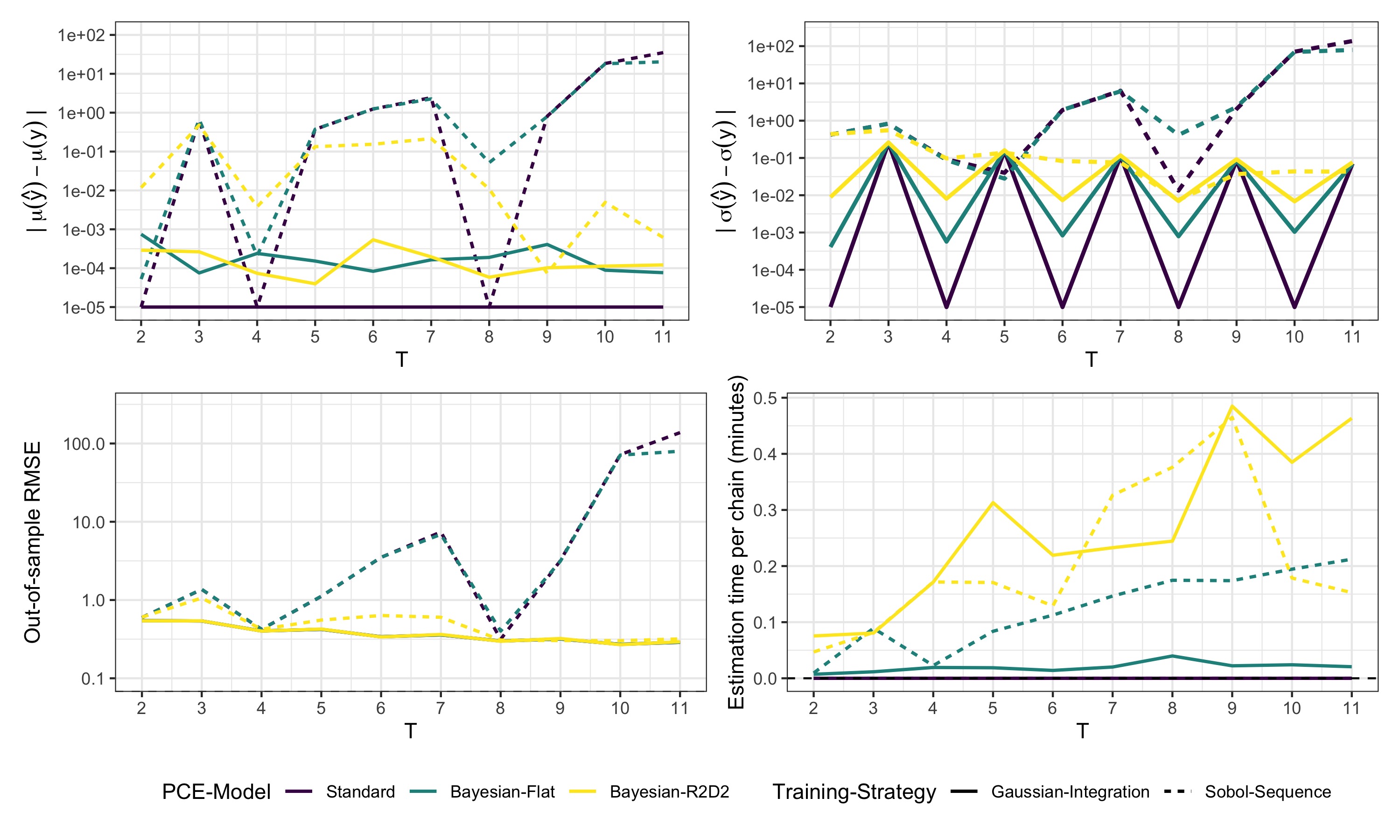}
    \caption{Summarized results for the Signum function by the number of training points $T$ (x-axis; log-scaled), type of PCE model of degree $d = T-1$ (colors), and strategy for choosing training points (line type). Top-left: absolute error of the mean $\mu$; top-right: absolute error of the standard deviation $\sigma$; bottom-left: average out-of-sample predictive RMSE; bottom-right: estimation times per fitted model. Smaller values are better. Absolute errors of mean and standard deviation are truncated from below at $1e-5$ to improve readability of the figure but actually go as far as $1e-14$.}
    \label{fig:signum-summaries-conv}
\end{figure}

A closer investigation of the approximation errors for varying degree $d$ (all with $T = M = d + 1$) is provided in Figure~\ref{fig:signum-summaries-conv}.
For the Sobol training points, we see almost uniformly higher accuracy of the Bayesian-R2D2 model as compared to the other two PCE approaches.
The superiority of Bayesian-R2D2 is particularly striking for the RMSE, but clearly visible for all evaluated error metrics.
For the training points given by Gaussian quadrature, all models perform very well overall as expected due to those training points' provable optimality. However, due to the sampling-based estimation procedure, both Bayesian approaches do not achieve the extremely high accuracy of Standard PCE in certain scenarios (absolute errors of mean and standard deviation below $1e-14$ in some cases). Increasing the number of MCMC samples $S$ would reduce the sampling error of the Bayesian approaches arbitrarily in theory, but scaling only by a factor of roughly $\sqrt{S}$, which renders errors below  $1e-4$ impractical to achieve (assuming unit scale of the target function). Whether the error is $1e-4$ or $1e-14$ is unlikely to matter in practice, but may still be of theoretical interest.

\newpage
\subsection{Carbon dioxide storage benchmark}
\label{CO2-case-study}

Finally, we consider a carbon dioxide storage benchmark \citep{koppel2019comparison, ik_data2017} 
for demonstration on a more realistic problem. This benchmark case describes the injection of carbon dioxide gas (CO$_2$) into a deep geological formation to mitigate greenhouse gas effects. Upon injection, the CO$_2$ spreads within the formation, which is initially filled with brine (i.e., with very salty, deep groundwater).

This test case allows to analyze the joint effects of several sources of uncertainty in the modelling of hydrosystems. In particular, these are uncertain boundary conditions (here: injection rate), uncertain material properties (here: porosity of the formation) and uncertainties in constitutive relations of multi-phase fluid flow (here: the degree to which the formation's permeability occurs in the so-called fractional flux function). It is assumed, that the density and viscosity of the fluids (CO$_2$, brine) are constant and that all processes are isothermal, such that CO$_2$ and brine can be described as a two-phase flow. When additionally assuming that the fluids' pressure can be determined in analytic form and upon transformation into radial coordinates, one can obtain the following governing equation:
\begin{equation}\label{eq:co2}
    \phi \frac{\partial S_g}{\partial t} 
    - \frac1r \frac{\partial}{\partial r} \left(q_{CO_2}C_p f_g \right) 
    -q_{CO_2} = 0,\quad \text{in}\; D\times(0,T_{\rm max}),
\end{equation}
which resembles a non-linear, hyperbolic partial differential equation for the (scalar) saturation of $CO_2$, denoted by $S_g$ (with subscript $g$ for gas). In this expression,
the uncertain porosity of the formation is denoted by $\phi$,
the time coordinate is denoted by $t\in (0,T_{\rm max})$,
the radial space location (coordinate) by $r \in D=[0,\,250]$,
the unknown saturation of the gas phase by $S_g$, and
the uncertain injection rate by $q_{CO_2}$.
Recall that the non-linear fractional flux function, here denoted by $f_g$, depends on the unknown $CO_2$ saturation $S_g$ and on the formation's permeability (symbols not shown here) to an uncertain degree. The constant $C_p$ represents the pressure decay.

The numerical solution of the non-linear equation \eqref{eq:co2} is obtained via a finite-volume method with central-upwind flux \citep{NUM:NUM20049}. The complete input and output data are available in \citep{ik_data2017}.

\begin{figure}
    \centering
    \includegraphics[width=.99\textwidth]{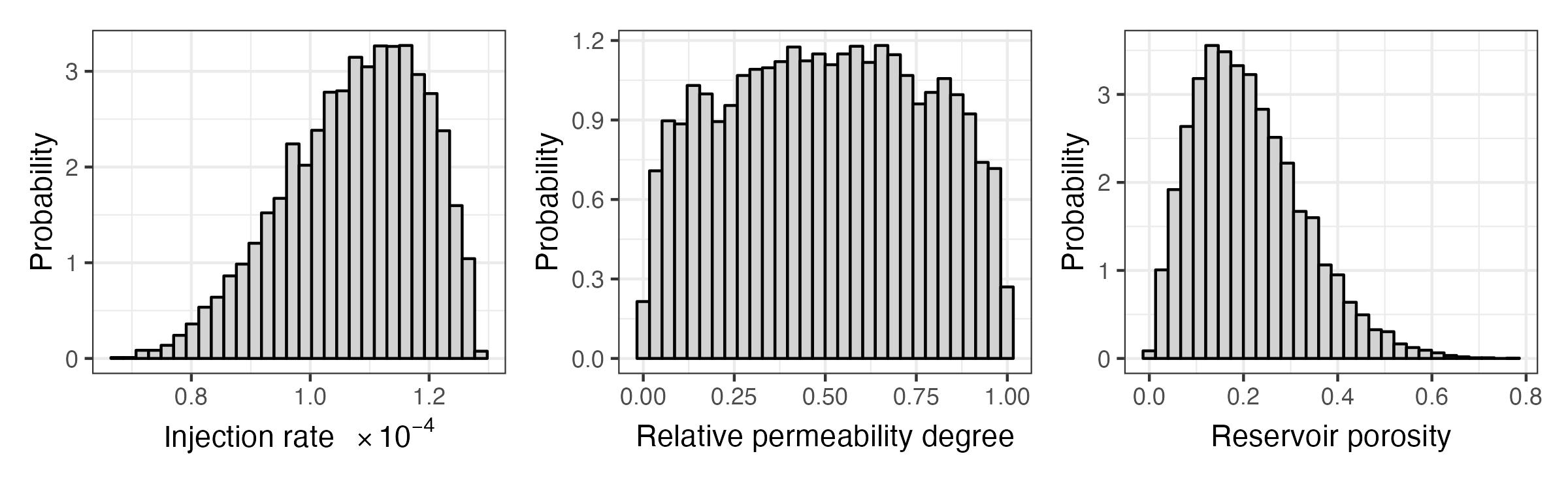}
    \caption{Marginal distributions of the input variables in the CO2 storage benchmark.}
    \label{fig:co2_sources}
\end{figure}


Following the original benchmark definition \citep{koppel2019comparison}, we analysed the CO$_2$ saturation data after $T_{\rm max}=100$ days at $L=250$ discretized spatial locations. We trained independent PCE models per coordinate on $T=(27,  64,  125,  216, 1331)$ training points given by $3$-dimensional grids of equidistant points across the $N=3$ input dimensions. The training data are available in the original benchmark definition \citep{koppel2019comparison}.

Due to strong discontinuities of the underlying physical model at certain combinations of parameter values and space/time coordinates, the approximation performance for PCE-based models is limited by design \citep{koppel2019comparison}. Hence, we cannot necessarily expect our proposed methods to perform competitively with methods specifically designed for such discontinuous target functions \citep{koppel2019comparison}. Rather, the goal is to show the performance achievable despite this limitation and compare the results to those obtained by other methods with the same limitation.  A nice property of this benchmarking case is that it has $L=250$ spatial locations, so that it actually represents $250$ test cases in one single scenario.

To obtain the polynomial basis for our PCE, we used the arbitrary polynomial chaos method (aPC) \citep{OladNowak_RESS2012} on the marginal distributions of input parameters as depicted in Figure~\ref{fig:co2_sources}. As aPC basis, we considered 3-dimensional aPC polynomials up to a total degree of $d = 10$, which implies $M = 286$ polynomials in total. As for the other case studies, these values were chosen to create a varying range of inference complexity. 

As full reference models (two for each value of $T$), we used all $M$ considered polynomials as basis, equipped with two different R2D2 priors on the $M - 1$ non-constant polynomials as described in Section \ref{sparse-bayesian-PCE}. The first R2D2 prior uses $R^2 \sim \text{Beta}(\zeta = 0.5, \nu = 2)$, while the second R2D2 prior uses $R^2 \sim \text{Beta}(\zeta = 0.9, \nu = 10)$ for reasons explained later.
Subsequently, with our two variable selection methods for comparison, we extracted the $M_{\rm sel} = (25, 50)$ most important out of all non-constant polynomials plus the constant polynomial. These selected polynomials were used as PCE bases for corresponding sparse aPC models. We investigated two different values for $M_{\rm sel}$, because the local discontinuities of the underlying true function were likely to require many polynomials to be approximated acceptably well. We used the same R2D2 priors for the sparse sub-models as for the corresponding full reference models. MCMC sampling methods were the same as for the other case studies. However, now we use only a single Markov chain per model to simplify parallelization across models on a computing cluster. Each chain was run for 3000 iterations, out of which the first 1000 were discarded as warmup. This provides a total of 2000 post-warmup draws used for inference for each model version. Once again, all models converged well as indicated by standard algorithm-specific and algorithm-agnostic convergence diagnostics.

As outcome measures, we considered the normalized $\ell^2$ norm of the mean and SD biases over the $L=250$ locations
\begin{align}
     \frac{\|\mu(\hat{y}) - \mu(y)\|}{L} &= \frac{\sqrt{\sum^{L}_{l=1} (\mu(\hat{y}_l) - \mu(y_l))^2}}{L} \\
     \frac{\|{\sigma(\hat{y})-\sigma(y)}\|}{L} &= \frac{\sqrt{\sum^{L}_{l=1} (\sigma(\hat{y}_l) - \sigma(y_l))^2}}{L}.
\end{align}
The variables $\mu(\hat{y})$ and $\sigma(\hat{y})$ are the length-$L$ vectors of model-implied mean and SD estimates, respectively. The vectors  $\mu(y)$ and $\sigma(y)$ are the corresponding vectors of "true" reference values obtained by \citep{koppel2019comparison} via Monte Carlo simulations using $10^4$ samples. These measures were chosen for comparability with the results obtained in \citep{koppel2019comparison}. Additionally, we considered the mean out-of-sample RMSE across locations
	\begin{equation}	
	\overline{\text{RMSE}} = \frac{1}{L}\sum^{L}_{l=1} \text{RMSE}_l,
	\end{equation}
where $\text{RMSE}_l$ is given by Equation~\ref{RMSE} applied to the $l$-th location.
	
\begin{figure}
    \centering
    \includegraphics[width=0.99\textwidth]{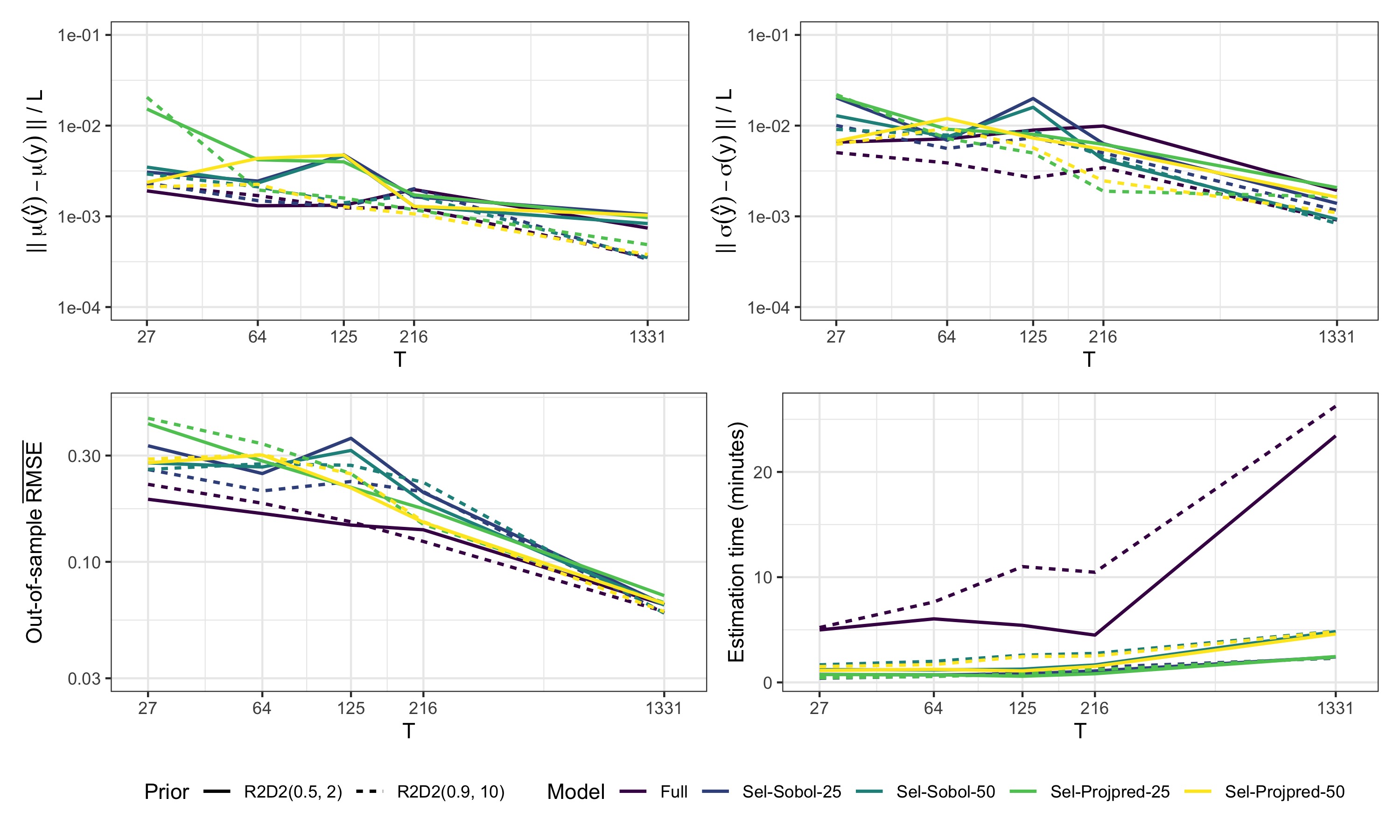}
    \caption{Summarized results for the CO2 data by the the number of training points (log-scaled; evaluated at $T=(27,  64,  125,  216, 1331)$), model type (colors), and prior on $R^2$ (line type). Top-left: normalized absolute error of the mean $\mu$; top-right: normalized absolute error of the standard deviation $\sigma$; bottom-left: average out-of-sample predictive RMSE; bottom-right: average estimation times per fitted model.}
    \label{fig:CO2-summaries}
\end{figure}

The summarized results across coordinates are depicted in Figure \ref{fig:CO2-summaries}. Especially for small to medium $T$, our PCE approaches are competitive with the state-of-the-art, but fall off a bit for larger $T$, where kernel greedy interpolation methods \citep{wendland2004, demarchi2005, wirtz2013} have an edge in this benchmark \citep[compare our results with those displayed in][Figure 7]{koppel2019comparison}. While not uniformly reaching the state-of-the-art on this benchmark due to the discussed discontinuities, our methods perform equally well or better than other PCE-based approaches \citep[compare with the blue lines displayed in][Figure 7]{koppel2019comparison} and provide better scaling potential to higher-dimensional problems and higher polynomial degrees. 

\begin{figure}
    \centering
    \includegraphics[width=0.99\textwidth]{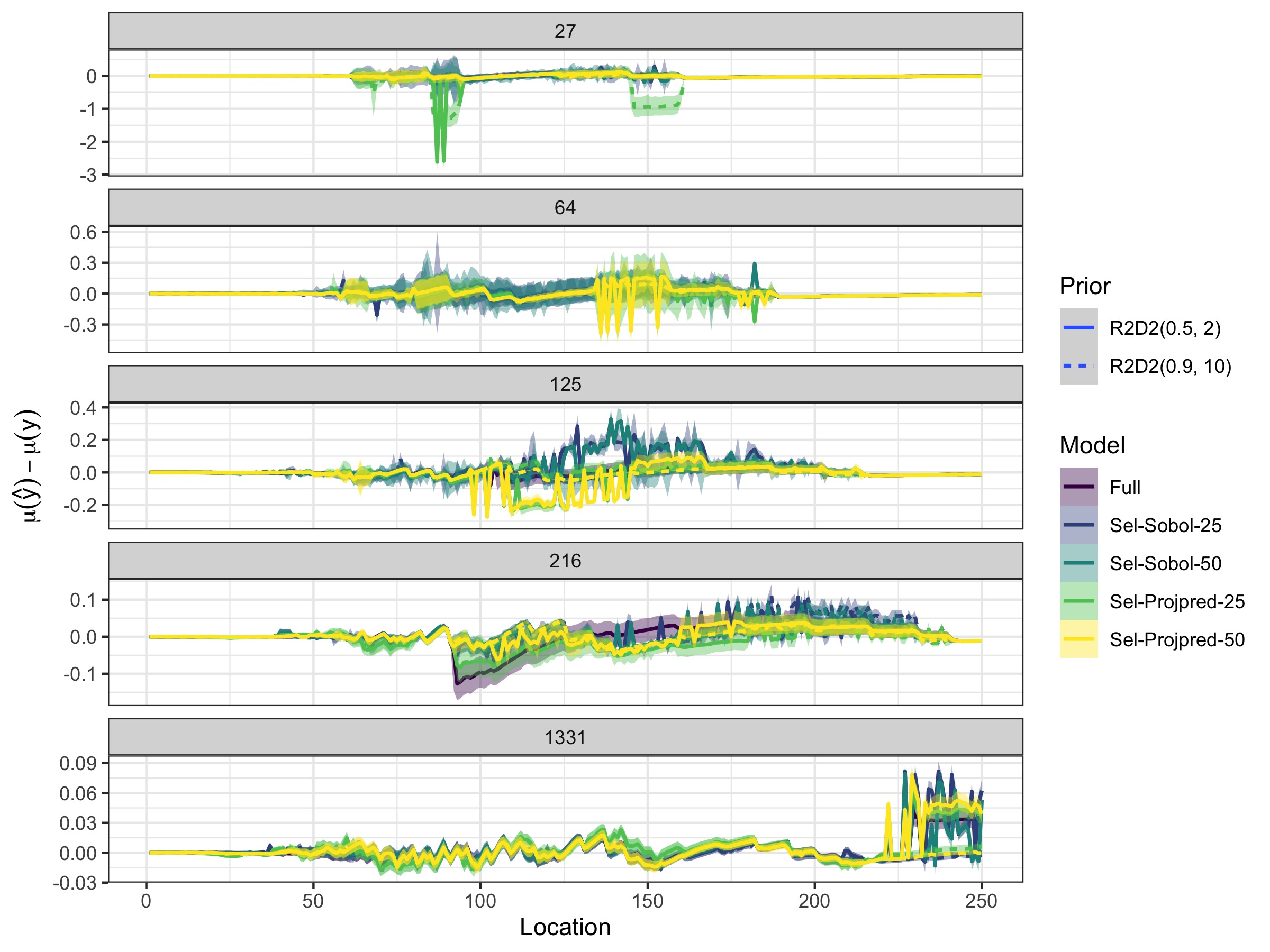}
    \caption{Difference between model-implied and true mean for the CO2 data by location (x-axis), number of training points $T$ (facets), and model type (colors). Values closer to zero are better.}
    \label{fig:CO2-diff-mean}
\end{figure}

\begin{figure}
    \centering
    \includegraphics[width=0.99\textwidth]{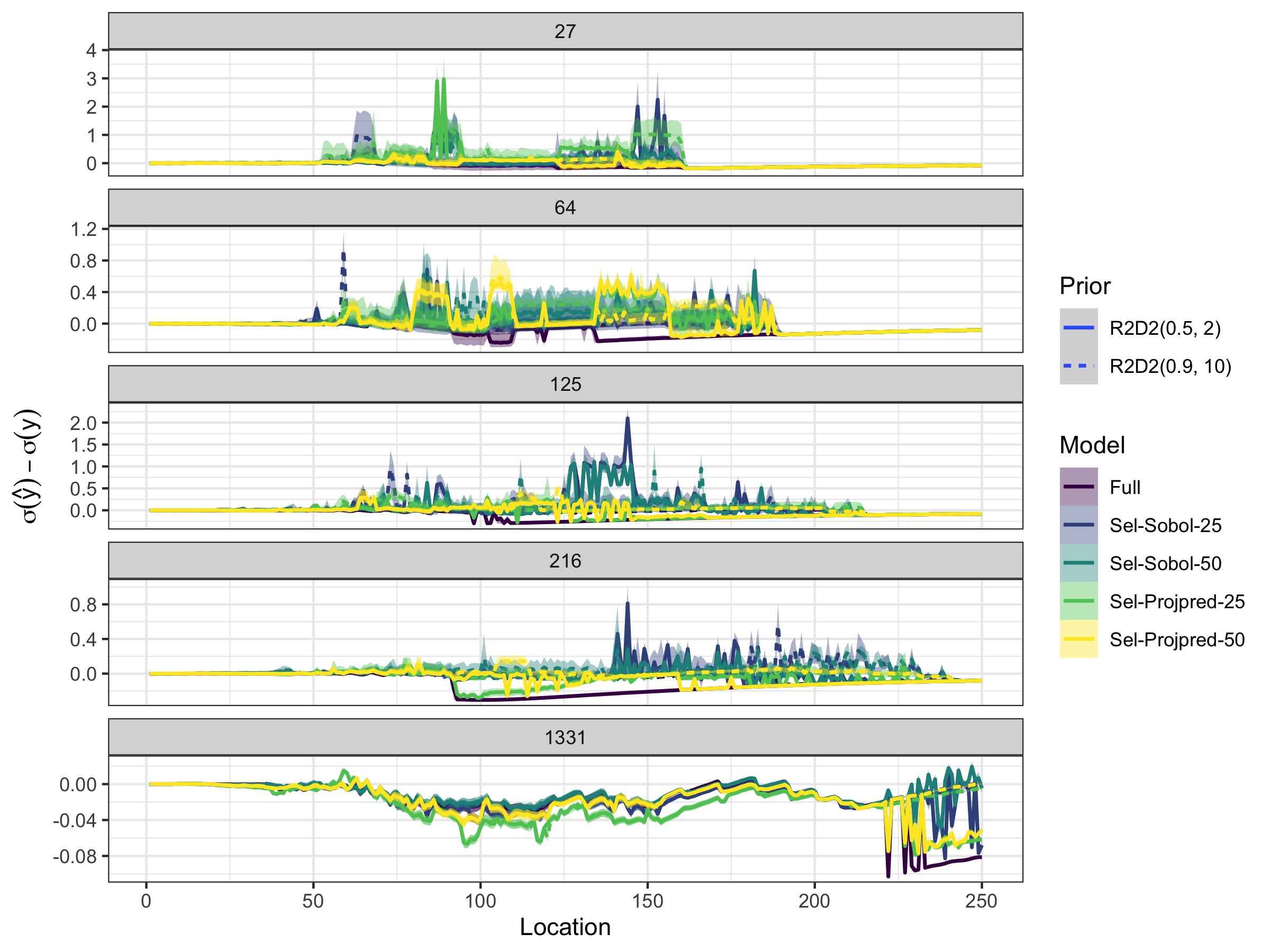}
    \caption{Difference between model-implied and true standard deviation for the CO2 data by location (x-axis), number of training points $T$ (facets), and model type (colors). Values closer to zero are better.}
    \label{fig:CO2-diff-sd}
\end{figure}

\begin{figure}
    \centering
    \includegraphics[width=0.99\textwidth]{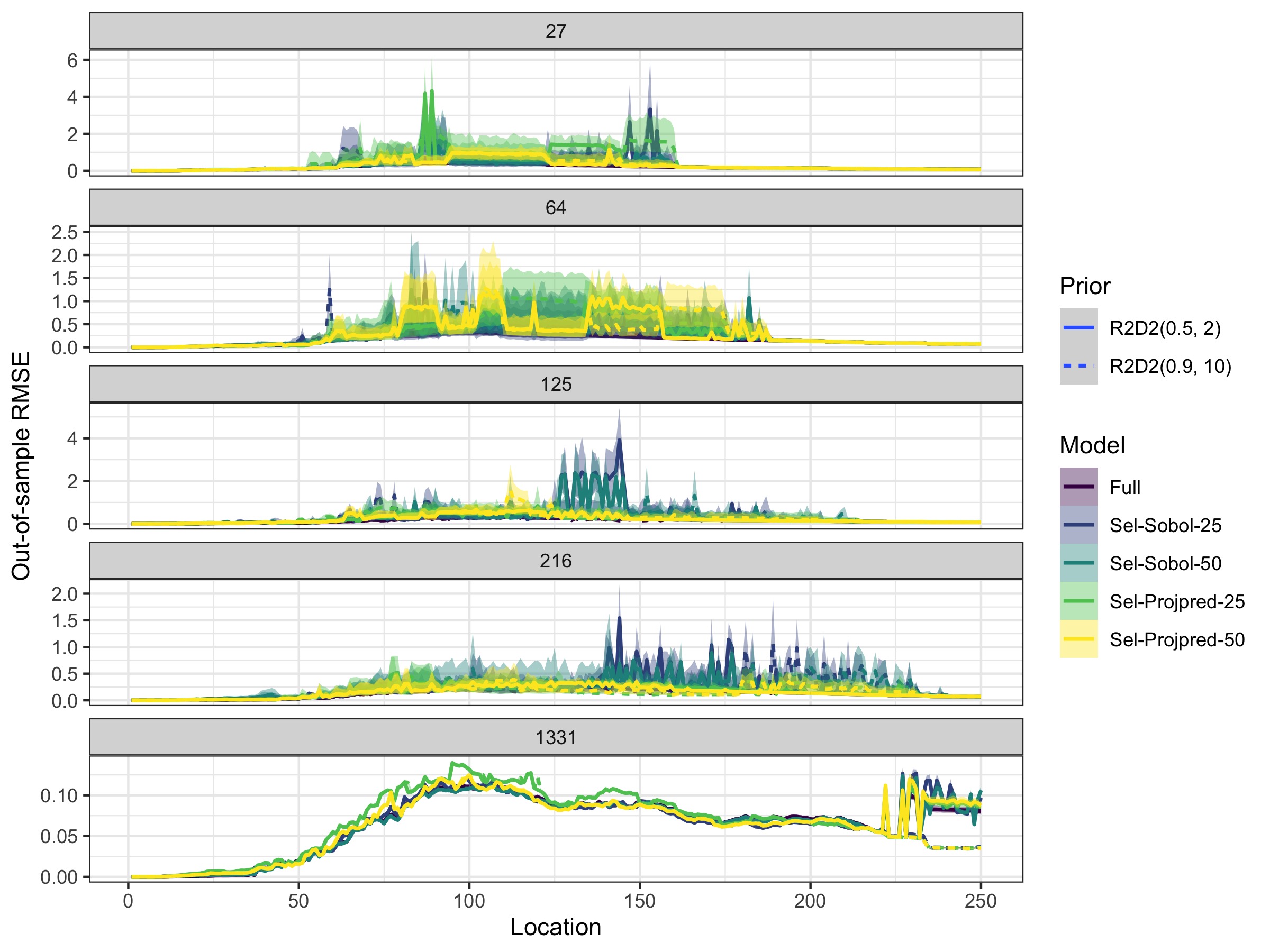}
    \caption{Out-of-sample predictive RMSE for the CO2 data by location (x-axis), number of training points $T$ (facets), and model type (colors). Values closer to zero are better.}
    \label{fig:CO2-RMSE}
\end{figure}

As can be readily seen, the full reference models show equal or better approximation performance than their respective sparse sub-models in terms of mean and SD bias as well as RMSE almost uniformly across $T$. Further inspection of the approximation performance for the individual locations, as displayed in Figures \ref{fig:CO2-diff-mean}, \ref{fig:CO2-diff-sd}, and \ref{fig:CO2-RMSE}, reveal that this is primarily due to approximation outliers for some coordinates $l \in [50, 200]$. These are the places where strong discontinuities and shocks are happening in the underlying true model. These outliers seem to occur less frequently for the sparse $M_{\rm sel} = 50$ models compared to the sparse $M_{\rm sel} = 25$ models, especially clearly so for projpred selection and small $T$. This suggests that many polynomials are required to provide an acceptable approximation of the true model for these locations, which could explain why the reference models perform better. Despite most of the $M=286$ polynomials being likely redundant, still including them in the model appears to do little harm in the considered scenarios. This points to the success of the R2D2 prior in shrinkage those redundant coefficient sufficiently to prevent overfitting.

Figures \ref{fig:CO2-diff-mean} and \ref{fig:CO2-diff-sd} reveal another problematic pattern that is present only for models with an $R^2 \sim \text{Beta}(\zeta = 0.5, \nu = 2)$ prior: For larger $T$ and specific locations ($l > 90$ for $T=218$ and $l > 220$ for $T=1331$) the mean and SD estimates of the reference model, and as a result also of the corresponding sparse sub-models, have substantial bias. This is because all PCE coefficients had been shrunken to almost zero, thus preventing both sensible prediction and coefficient selection. Similar problems have been observed for other PCE-based approaches applied to this CO$_2$ data set \citep[see Figure 3 in][]{koppel2019comparison}. Further inspection revealed that this behavior occurred in our models because of a bimodal posterior of the PCE coefficients, with one mode at zero for all coefficients and the other "sensible" mode away from zero for at least some of the truly relevant coefficients.

In additional experiments not shown here, we were able to find this second mode at least for $T=1331 > M$ by setting the initial values of the MCMC chains to the maximum-likelihood estimate (MLE). In this case, the resulting mean and SD estimates of the Bayesian models were almost unbiased again, closely resembling the corresponding values of the models with an $R^2 \sim \text{Beta}(\zeta = 0.9, \nu = 10)$, which implies a weaker R2D2 prior in the sense that its shrinkage effect is reduced (expected $R^2$ is $\zeta = 0.9$ instead of $\zeta = 0.5$). However, since the MLE does not exist for $T < M$, it is non-trivial to find such suitable initial values in general. Instead, one either has to apply a weaker R2D2 prior, as done here,  to practically eliminate the all-zero mode; or one has to run multiple MCMC chains with diffuse initial values in the hope that at least one finds the desired mode.

In summary, the results demonstrate that our sparse Bayesian PCE can perform well even when approximating discontinuous target functions that are not well approximated by polynomials in general. However, care must be taken that the applied shrinkage priors do not "over-shrink" all PCE coefficients to zero (thus preventing any signal to be encoded in the PCE) and that enough polynomials are still included in the sparse representations.

\section{Discussion}

In this paper, we presented a fully Bayesian approach to sparse polynomial chaos expansion (PCE). In particular, we define the PCE coefficients as random variables to be inferred probabilistically based on training data (e.g., simulations performed by a computationally expensive model for which one wants to build a surrogate). The desired sparsity is induced by a specific choice of prior over the PCE coefficients, known as the R2D2 prior \cite{zhang_bayesian_2020}. Below we recap and summarize our methods and obtained results. 

The R2D2 prior belongs to the class of priors known as shrinkage priors in statistics. It consists of two parts: (1) It regularizes the proportion of variance in the true responses (e.g., of the results of the expensive model) that is captured by the PCE, that is, it regularizes the coefficient of determination $R^2$ as known from statistical regression. (2) It controls how the captured variance is distributed over the expansion coefficients. If desired, the latter could be informed by assumptions on how the PCE spectrum decays with increasing polynomial degree. As the variance captured by individual PCE terms is given by the square of the corresponding coefficient (and the overall captured variance is the sum of squares, without the zeroth-degree term), both parts of the R2D2 prior link naturally to PCE theory. 

The R2D2 induces only weak sparsity in the sense that, upon training, many coefficients are \emph{close} to zero, but not necessarily exactly zero. Thus, we post-process the weakly sparse PCE with a procedure of statistical variable selection to obtain truly sparse expansions. Finally, we re-train these truly sparse expansions with the same training data, and again with the same R2D2 prior; thus, we obtain the final, sparse expansion. For variable selection, we compare a well-known Sobol-based, greedy variable selection and a more recent, promising variable selection method called projective prediction \cite{piironen_projective_2020}. The latter resembles a more global view on the sparse selection of PCE terms, and is linked to Bayesian cross-validation.

In combination, the R2D2 prior and the variable selection step allow for a global search for sparse PCEs. The global search can consider all candidate terms defined by a high-degree PCE, based on rigorously statistical tools. This goes far beyond current, rather heuristic, search strategies for relevant PCE terms, which mostly search only in the vicinity of already identified terms. Also, our approach allows to decouple the computational design of training points (i.e., via collocation or integration rules) from the search for sparse PCEs: it can be applied even on given sets of training data, and can search through a space of possible PCE expansions that even includes strongly under-determined cases. The R2D2 prior not only enables sparse selection afterwards, but also directly regularizes against overfitting and oscillations, such that sparse selection may not even be required necessarily.

We demonstrated these favourable properties in four different test cases. The test cases included the Signum function to assess how our approach handles the Gibbs and Runge phenomena of oscillation in polynomial approximations. We also used the Ishigami function and the Sobol function; these are known, hard cases for benchmarking PCE techniques. These three functions all contain known patterns of sparsity (e.g., only even or only odd degree, independent variables without cross terms, or sets of close-to-insignificant variables). In all these cases, our sparse Bayesian PCE with R2D2 prior performed very well. For the Signum function, it was equally good as the standard PCE when optimal training points are used, and it outperformed the standard PCE for suboptimal points, as the R2D2 reduces oscillations.

For exhaustive comparison to many other surrogate methods (including different versions of the PCE), the fourth test case was a benchmark on CO\textsubscript{2} storage in geological formations. On this test case, we compared the results and performance of our sparse Bayesian PCE to an extensive list of benchmarked surrogate techniques published in a comparison paper \cite{koppel2019comparison}. Here, we could observe that our method outperforms existing polynomial-type surrogate methods. In all our test cases, the projective prediction method for sparse selection outperformed the Sobol-based selection.

Care must be taken in the specific hyperparameter choice of the R2D2 prior. In our study, we explored different Beta priors for $R^2$, and found that the choice of mean and variance in the Beta distribution may require some expert knowledge or tuning (see also \citep{zhang_bayesian_2020, aguilar_R2D2M2_2022}). With the R2D2 prior, the PCE coefficients are no longer found by analytically solving a linear system, but by sampling via Markov Chain Monte-Carlo methods. This is much more expensive computationally, but also provides information about the uncertainty of the inferred PCE without making any distributional assumptions of the posterior. 

Detailed analyses of the computational costs revealed that they are indeed mainly driven by MCMC sampling of the full model that contains all PCE polynomials up to a given degree. This took around 1-3 minutes in simple cases to around 3 hours in the most complicated case we evaluated (i.e., Sobol function with 8100 training points and 3003 polynomials). Running MCMC again for the sparse sub-models took no more than some seconds in
all cases due to the small number of remaining model parameters. In comparison, constructing the PCE polynomials themselves took only a fraction of a second in all cases and constructing the search path via projpred took less than a second in simple cases up to around 30 seconds in the most complex case.

Whether the effort of MCMC sampling is worth the effort depends primarily on the computational complexity of the original simulator that produces the training data. If simulations are fast (i.e.,
seconds or minutes per simulation), then we do not have to bother with
any sparse PCE approach at all and can just use standard (unregularized)
PCE or any other surrogate model of our liking. However, when each
simulation takes many hours or even days (which is not uncommon in physics applications; e.g., \cite{Koeppel2017}), then
we would rather invest a few minutes or hours for our fully Bayesian sparse
PCE approach than to wait many days for sufficient additional
simulations to finish before standard PCE (or comparable approaches)
would achieve similar accuracy.


In terms of scaling, our proposed method behaves like other basis-function approximations, whether those are PCEs, wavelets, splines, or basis-function approximations to Gaussian processes \citep{james_introduction_2013, wood_gams_2017, riutort-mayol_practical_2022}. That is, their computational costs and memory requirements scale well (linearly) with the number of training points, but badly (exponentially) with the number of input dimensions due to the use of tensor product terms (see also Section \ref{general-PCE}). Because of the latter, we would not recommend using full PCE -- or any other basis function approximation -- for more than about 4 input dimensions (see also \citep{riutort-mayol_practical_2022}). In these cases, other surrogate modeling approaches such as exact Gaussian processes or deep neural networks might be favorable. Yet, enforcing sparsity in problems that are rather sparse by nature can postpone the effect of the curse of dimensionality towards higher dimensions.

The idea of sparse construction of basis function approximations by means of Bayesian linear models equipped with the R2D2 prior and subsequent variable selection is quite general. We applied it to the arbitrary polynomial chaos expansion, which is not bound to specific cases for probability distributions of model input parameters. In future research, our approach could be applied to adaptive multi-element PCE versions, could be equipped with schemes for active learning, or could be combined with Gaussian processes for handling the residuals of the found sparse PCEs.


\section*{Acknowledgments}

Partially funded by Deutsche Forschungsgemeinschaft (DFG, German Research Foundation) under Germany’s Excellence Strategy - EXC 2075 – 390740016 and DFG Project 432343452.

\section*{Appendix}

\subsection*{Appendix A: Beta and Dirichlet distributions}

Below, we show densities and the first two moments of the Beta and Dirichlet distributions. 
The density of the Beta distribution for scalar $x \in [0,1]$ with positive shape parameters $a_1$ and $a_2$ is given by
\begin{equation}
    p_{\rm Beta}(x \,|\, a_1, a_2) = \frac{x^{a_1-1} (1-x)^{a_2-1}}{B(a_1, a_2)},
\end{equation}
where $B(a_1, a_2)$ is the two-dimensional Beta function. In this study, we are using the parameterization of the Beta distribution that arises when we set $a_1 = \zeta \nu$ and $a_2 = (1-\zeta) \nu$. In this so-called "mean-precision" parameterization, the mean and variance of the Beta distribution are given by
\begin{equation}
    \mathbb{E}(x) = \zeta \quad \text{ and } \quad \text{Var}(x) = \frac{\zeta (1-\zeta)}{1+\nu}.
\end{equation}

The density of the Dirichlet distribution for a simplex $x \in [0,1]^M$ and $\sum_{i=1}^M x_i = 1$ with concentration parameter vector $\theta \in \mathbb{R}_+^M$ is given by
\begin{equation}
    p_{\rm Dirichlet}(x \,|\, \theta) = \frac{1}{B(\theta)} \prod_{i=1}^M x^{\theta-1},
\end{equation}
where $B(\theta)$ is the $M$-variate Beta function. When we set $\theta_0 = \sum_{i=1}^M \theta_i$, the mean and variance of the Dirichlet distribution for the $i$th component $x_i$ are given by
\begin{equation}
    \mathbb{E}(x) = \frac{\theta_i}{\theta_0} \text{ and } \quad \text{Var}(x) = \frac{\theta_i / \theta_0 \times (1 - \theta_i / \theta_0)}{1+\theta_0}.
\end{equation}

\subsection*{Appendix B: Robustness to noisy training data}

\begin{figure}
    \centering
    \includegraphics[width=0.99\textwidth]{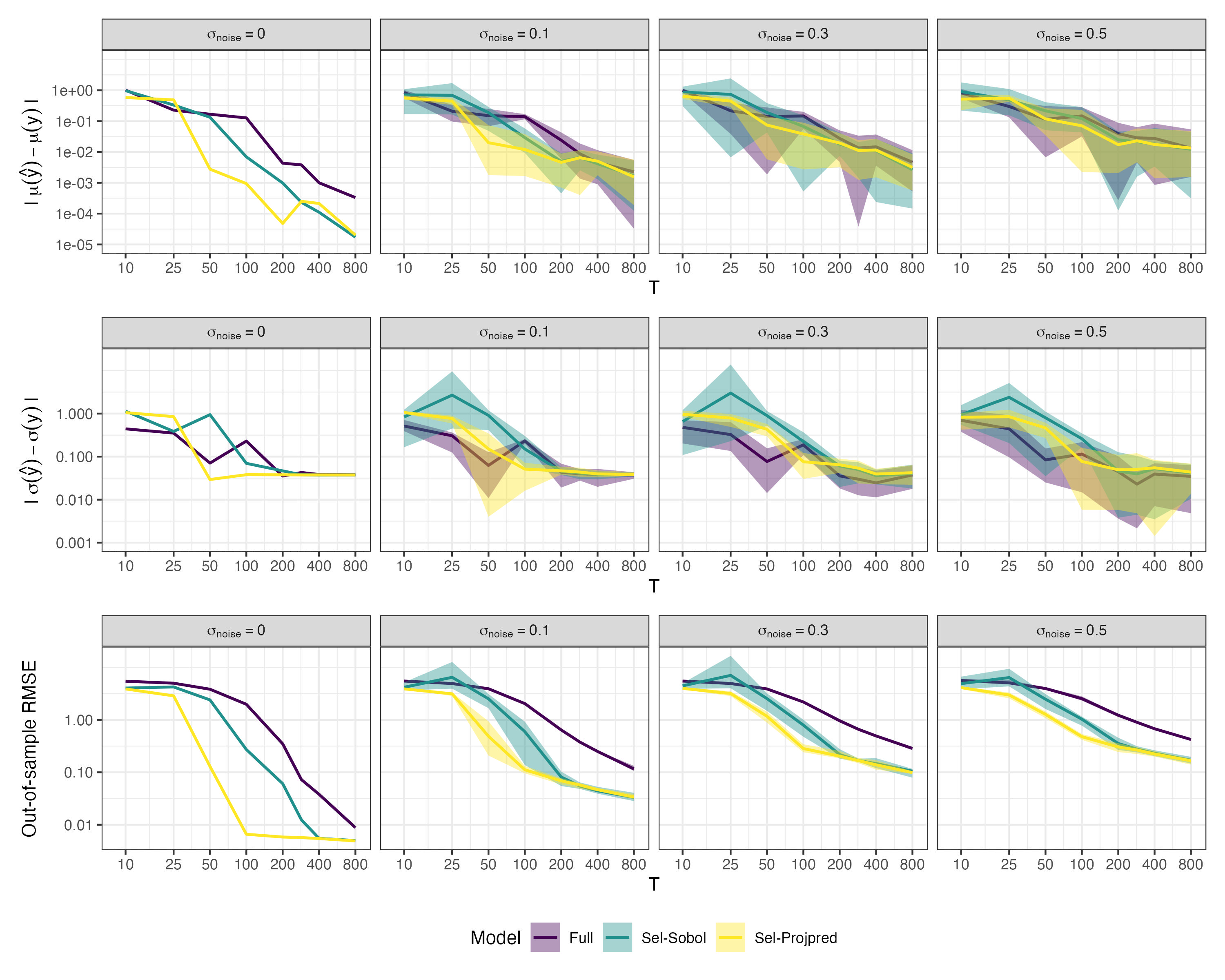}
    \caption{Summarized results for the Ishigami function $(a=7, b=0.1)$ by the number of training points $T$ (log-scaled; evaluated sizes are equal to the displayed axis ticks), model type (colors), and standard deviation of noise (facets). Top: absolute error of the mean $\mu$; middle: absolute error of the standard deviation $\sigma$; bottom: average out-of-sample predictive RMSE.}  
    \label{fig:ishigami-summaries-noise}
\end{figure}

In order to test the robustness of our method against noise in the training data, we re-ran our experiments of the Ishigami function from Section~\ref{ishigami-case-study} with varying amounts of noise. Specifically, we added noise from a ${\rm normal}(0, \sigma_{\rm noise})$ distribution where the standard deviation took on values of $\sigma_{\rm noise} = (0.1, 0.3, 0.5)$. To reduce dependency on individual noisy simulations, we created 10 simulated training data sets per condition. The results depicted in Figure~\ref{fig:ishigami-summaries-noise} demonstrate the relative robustness of both the full reference model and the sparse projpred model against noise. This is in line with previous research that demonstrated that strongly regularized reference models act as a noise filter that reduces overfitting and that projpred variable selection is an especially powerful reference model approach \citep{pavone_reference_2020}. While the results show roughly the same overall patterns and orderings of the methods as the noiseless experiments, having added noise of first-digit order, accuracy in the third digit or lower becomes hard to achieve. Interestingly, the sparse sub-models resulting from greedy selection of the largest Sobol coefficients show unstable approximation behavior for $T=25$ training points for all degrees of added noise. That is, for some of the noisy data sets, the error metrics were multiple times worse than for the other methods. This adds more evidence that greedy selection of terms may be a sub-optimal and potentially unstable approach even if selection was done based on a well-performing reference model.

\newpage
\bibliographystyle{acm}
\bibliography{references}

\end{document}